\PassOptionsToPackage{table,xcdraw}{xcolor}
\documentclass[lettersize,journal]{IEEEtran}
\usepackage{amsmath,amsfonts}
\usepackage{algorithmic}
\usepackage{algorithm}
\usepackage{array}
\usepackage[caption=false,font=normalsize,labelfont=sf,textfont=sf]{subfig}
\usepackage{textcomp}
\usepackage{stfloats}
\usepackage{url}
\usepackage{verbatim}
\usepackage{graphicx}
\usepackage[
  backend=biber,
  style=ieee,
  maxnames=999,
  maxbibnames=999
]{biblatex}
\usepackage{bm}
\usepackage{pgf}

\usepackage[switch,pagewise]{lineno}

\usepackage{hyperref}
\usepackage{gensymb} %
\usepackage{color,soul} %
\usepackage{siunitx} %
\usepackage{cleveref}
\usepackage{booktabs}
\usepackage{multirow}
\usepackage{float}
\usepackage{pifont}
\usepackage{amssymb}
\usepackage{adjustbox} %
\usepackage{threeparttable} %
\usepackage{hhline} %
\usepackage{xcolor} %

\newcommand{\cmark}{\ding{51}}%
\newcommand{\xmark}{\ding{55}}%

\usepackage{epsf,caption,subcaption}
\usepackage{amsmath}

\makeatletter
\g@addto@macro\@eqnnum{\aftergroup\noindent}
\makeatother

\newcommand{\Figref}[1]{\hyperref[#1]{Figure~\ref{#1}}}
\newcommand{\figref}[1]{\hyperref[#1]{Fig.~\ref{#1}}}

\AtNextBibliography{\footnotesize} %
\AtEveryBibitem{\clearfield{doi}} %
\AtEveryBibitem{\clearfield{url}\clearfield{eprintclass}} %
\AtEveryBibitem{\clearfield{urldate}}
\AtEveryBibitem{\clearfield{editor}} %
\DeclareFieldFormat[misc]{title}{``#1''} %
\DeclareFieldFormat{eprint:arxiv}{\mkbibemph{arXiv:#1}} %
\DeclareFieldFormat[online]{title}{\mkbibemph{#1}} %
\usepackage[font=footnotesize,labelsep=period]{caption}

\newcommand{\red}[1]{#1}

\begin{document}

\newcommand{\eg}{\textit{e.g.},~}
\newcommand{\etal}{\textit{et~al.}}

\newcommand{\omnfwayoshot}{96.8 ± 1.6\%} %
\newcommand{\omnfwayfshot}{98.8 ± 0.5\%} %
\newcommand{\omntwayoshot}{89.1 ± 1.3\%} %
\newcommand{\omntwayfshot}{96.1 ± 0.5\%} %
\newcommand{\omnthwayoshot}{83.3 ± 1.2\%} %

\newcommand{\hltodo}[1]{#1}

\definecolor{bred}{RGB}{192, 0, 0}

\title{Chameleon: A \red{Multiplier-Free} Temporal Convolutional Network Accelerator for End-to-End Few-Shot and Continual Learning\\from Sequential Data

}

\author{Douwe den Blanken,
\IEEEmembership{Graduate Student Member, IEEE},
and
Charlotte Frenkel,
\IEEEmembership{Member, IEEE}

\thanks{Douwe den Blanken and Charlotte Frenkel are with the Microelectronics Department (EEMCS Faculty), Delft University of Technology, 2628 CD Delft, Netherlands (e-mail: d.m.j.denblanken@tudelft.nl; c.frenkel@tudelft.nl).

This work has been published in the \textit{IEEE Journal of Solid-State Circuits (JSSC)}, Digital Object Identifier (DOI): 10.1109/JSSC.2025.3645640.

\copyright~2026 The Authors. This work is licensed under a Creative Commons Attribution 4.0 License, see \url{https://creativecommons.org/licenses/by/4.0/}.}
}

\maketitle

\begin{abstract}

On-device learning at the edge enables low-latency, private personalization with improved long-term robustness and reduced maintenance costs. Yet, achieving scalable, low-power end-to-end on-chip learning, especially from real-world sequential data with a limited number of examples, is an open challenge. Indeed, accelerators supporting error backpropagation optimize for learning performance at the expense of inference efficiency, while simplified learning algorithms often fail to reach acceptable accuracy targets. In this work, we present Chameleon, leveraging three key contributions to solve these challenges. (i)~A unified learning and inference architecture supports few-shot learning (FSL), continual learning (CL) and inference at only 0.5\% area overhead to the inference logic. (ii)~Long temporal dependencies are efficiently captured with temporal convolutional networks (TCNs), enabling the first demonstration of end-to-end on-chip FSL and CL on sequential data and inference on 16-kHz raw audio. (iii)~A dual-mode, \red{multiplier-free} compute array allows either matching the power consumption of state-of-the-art inference-only keyword spotting (KWS) accelerators or enabling $4.3\times$ higher peak GOPS. Fabricated in 40-nm CMOS, Chameleon sets new accuracy records on Omniglot for end-to-end on-chip FSL (96.8\%, 5-way 1-shot, 98.8\%, 5-way 5-shot) and CL (82.2\% final accuracy for learning 250 classes with 10 shots), while maintaining an inference accuracy of 93.3\% on the 12-class Google Speech Commands dataset at an extreme-edge power budget of \qty{3.1}{\micro\watt}.%
\end{abstract}


\vspace{-12pt} %

\section{Introduction}

\begin{figure}[!t]
\centering
\includegraphics[width=0.46\textwidth,trim=20 31 23 39, clip]{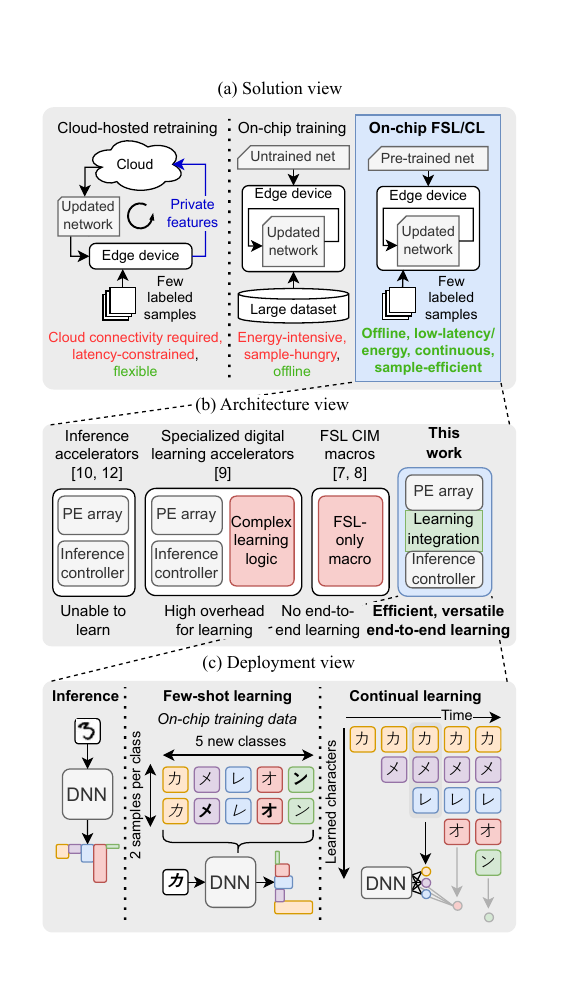}
\caption{(a)~Overview of three approaches to endow edge AI devices with the ability to adapt, where on-device few-shot learning (FSL) and continual learning (CL) enable offline, low-latency, and sample-efficient learning. (b)~Comparison of an inference and three learning architectures, including ours; we minimize the footprint of learning by integrating it into the inference pipeline rather than treating it separately. (c)~Our architecture supports end-to-end on-chip inference, FSL, and CL at a low overhead.}
\label{fig:key_ideas}
\end{figure}

\IEEEPARstart{T}{he} rise of edge computing has driven demand for deploying deep learning models on resource-constrained devices \cite{l2l_on_chip_marian_reptile}, forming the backbone of the Internet-of-Things~(IoT) ecosystem \cite{ravaglia2021tinyml}. However, most edge artificial intelligence~(AI) devices focus on inference \cite{ravaglia2021tinyml}, relying on pre-trained models that cannot be adapted post-deployment. This lack of adaptability limits robustness to emerging features and data distribution shifts \cite{ravaglia2021tinyml}, making smart sensors unreliable over time \cite{amodei_ai_safety_problems}. \autoref{fig:key_ideas}(a) shows that while cloud-hosted retraining offers a workaround where new labeled data is uploaded from an edge device to a central server for model updates, this approach is cloud-connectivity-dependent and does not fit latency-constrained scenarios, while leading to privacy concerns \cite{pellegrini2020latent_reply}. On the other hand, training a model from scratch on-device incurs high energy penalties and still requires the availability of a large labeled dataset, which is not realistic at the edge. Therefore, enabling efficient on-device learning at the edge requires the ability to continually learn from limited data. %

\vfill

\IEEEpubidadjcol

However, over a device's lifetime, far fewer tasks are learned than inferences performed. Hence, it is key for an edge AI device to support learning without degrading inference performance and efficiency (\autoref{fig:key_ideas}(b)). Yet, current efforts to extend inference accelerators with few-shot learning (FSL) capabilities incur significant area, energy and accuracy penalties. For example, using quantized gradient descent (GD) to implement on-chip FSL \cite{li2021quantization_maml, l2l_on_chip_marian_reptile, exploring_quant_in_fsl_maml}, requires maintaining all intermediate activations in memory, transposing weight matrices, and using costly data types such as floating-point (FP) representations \cite{l2l_on_chip_marian_reptile}. %
Alternatively, gradient-free FSL approaches using compute-in-memory (CIM) macros~\cite{other_fsl_cim, sapiens} have been proposed to accelerate distance computation between input and stored embedding vectors for FSL. However, they fail to provide \textit{end-to-end} learning support as they rely on off-chip FP32 deep neural networks (DNNs) to compute high-quality embeddings. To the best of our knowledge, only one prior work computes the DNN embeddings on chip for gradient-free FSL~\cite{fslhdnn}. This design adds dedicated learning hardware alongside inference modules, yielding a $2\times$ increase in peak power, a 25-\% area overhead, and requires external memory to store all model parameters for learning, undermining its suitability for extreme edge scenarios.

Another challenge is that state-of-the-art (SotA) on-chip FSL architectures lack the ability to handle temporal information in two key ways. First, they cannot process \textit{sequential data}. All existing FSL accelerators target image classification tasks \cite{li2021quantization_maml, l2l_on_chip_marian_reptile, exploring_quant_in_fsl_maml, l2l_on_chip_marian_reptile,other_fsl_cim, sapiens,fslhdnn}, whereas real-world applications often involve sequential data, such as keyword spotting (KWS) where audio is typically sampled at frequencies ranging from \qtyrange{16}{48}{\kilo\hertz}. However, due to the limited input context length of current end-to-end KWS accelerators (60 \cite{vocell, giraldo_efficient_2021, vikram_tinyvers} to 101 \cite{bernardo_ultratrail_2020}), feature extraction (FE) such as mel-frequency cepstral coefficients (MFCC) is typically performed for two-order-of-magnitude reductions in input sequence length, at the expense of a 50--75\% area overhead \cite{vocell,kwantae_kim_23_uw_ring_oscillator_ro_kws}. Hence, capturing temporal dependencies that span multiple orders of magnitude is not only key for complex temporal tasks, it also allows streamlining system design. Second, current on-chip FSL designs cannot perform \textit{sequential learning of tasks with data distributions that shift over time}. This \textit{continual learning} (CL) scenario is typical of edge applications, where users add features, such as new keywords, over time. While high-quality embeddings produced by off-chip embedders have recently been shown to also support CL over up to 100 classes \cite{karunaratne2022memory_continual_learning_hermes_cl}, end-to-end on-chip CL is still an open challenge.

Overall, due to limited on-chip memory and strict power, area, and energy budgets, no extreme-edge solution has been proposed that effectively balances efficiency and versatility for end-to-end learning on sequential data. In this work, we present Chameleon, the first work to demonstrate scalable on-chip FSL and CL on sequential data, in a fully end-to-end fashion, and without degrading the inference performance or efficiency (\autoref{fig:key_ideas}(c)). We achieve this through three key contributions:

\begin{enumerate}
    \item To minimize the footprint of on-device learning, we build on \textit{prototypical networks} (PNs)~\cite{snell2017prototypical} to reframe FSL using distance computation as a forward pass through an equivalent fully connected (FC) layer, thereby leading to a \textit{unified learning and inference architecture}. %
    By integrating learning within the inference process, we enable gradient-free, end-to-end FSL and CL at only 0.5\% area overhead and 0.02\% added latency while setting new FSL accuracy records (96.8\%, 5-way 1-shot) and CL accuracy (82.2\% final, 89.0\% average for learning 250 classes with 10 shots) on the Omniglot~\cite{lake_omniglot} dataset.
    \item To enable accurate FSL and CL with temporal information, a DNN that captures long-range relationships to produce high-quality embeddings is crucial. Hence, we propose to use temporal convolutional networks (TCNs), whose parameter efficiency scales better than recurrent neural networks~\cite{bai2018empirical}, as efficient on-chip PN embedders for sequential data at the edge. Compared to SotA TCN accelerators~\cite{scherer2022tcn_cutie_benini,giraldo_efficient_2021,bernardo_ultratrail_2020}, we expand the receptive field by $160\times$ while cutting activation memory by $4\times$.
    \item To enable inference at an extreme-edge power budget while facilitating a high throughput for learning, we introduce a dual-mode \red{multiplier-free} processing element (PE) array \red{employing only bit shifters}. Combined with system-level power gating, we achieve \SI{3.1}{\micro\watt} real-time KWS or $4.3\times$ higher peak GOPS than SotA KWS accelerators.
\end{enumerate}
The remainder of this article is structured as follows: \autoref{sec:background} covers DNN training for FSL and silicon deployments for various FSL methods, while the proposed architecture of the Chameleon system-on-chip (SoC) is introduced in \autoref{sec:soc_architecture}. \autoref{sec:measurements} then presents the measurement results for FSL, CL and inference tasks, with concluding remarks in \autoref{sec:conclusion}. To promote reproducibility, reuse, and improvement, all code for this paper is open-source, including the accelerator source code, the training framework, and the test/simulation setup.\footnote{\url{https://github.com/cogsys-tudelft/chameleon}}

\section{Background}
\label{sec:background}

To give more insight into FSL algorithms, \autoref{sec:fsl_formalized} presents the \textit{meta-learning} framework used to train DNNs for FSL \cite{maml_finn}, also known as \textit{learning to learn}. \autoref{sec:fsl_on_chip_suitability} then examines various meta-learning methods and their suitability for on-chip deployment.

\subsection{Meta-Learning Framework}
\label{sec:fsl_formalized}

\begin{figure}[t]
\centering
\includegraphics[width=0.45\textwidth,trim=0 35 0 15, clip]{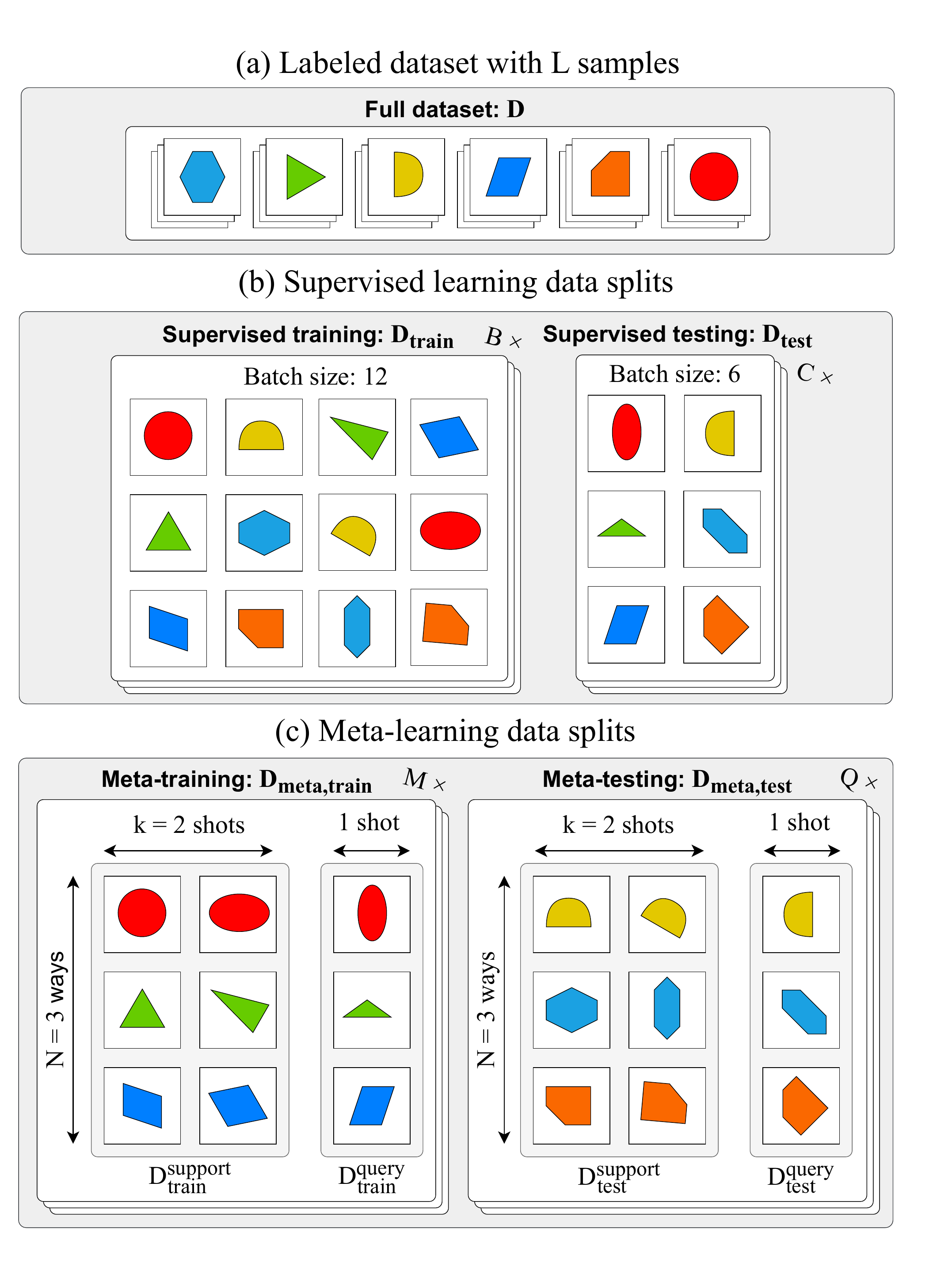}
\caption{(a)~Standard dataset comprising $L$ labeled examples across multiple classes. (b)~Supervised training and testing splits, illustrated for batch sizes of 12 and 6, respectively. (c)~Meta-training and meta-testing splits for an example 3-way 2-shot ($D^{\text{support}}$) task with 1-shot query data ($D^{\text{query}}$). Notice that the classes do not overlap between $D_{\text{meta,train}}$ and $D_{\text{meta,test}}$.}
\label{fig:meta_learning_data}
\end{figure}

In supervised learning, given a standard labeled dataset $D=\left\{\left(x_1,y_1\right),...,\left(x_L,y_L\right)\right\}$ with $L$ examples $x_i$ and labels $y_i$ (\autoref{fig:meta_learning_data}(a)), the training $D_{\text{train}}$ and testing $D_{\text{test}}$ splits share classes but not examples (\autoref{fig:meta_learning_data}(b)). However, in meta-learning, entire classes are assigned to training or testing splits (\autoref{fig:meta_learning_data}(c)), ensuring each split has a distinct set of classes.

For meta-training, a set of $M$ tasks \cite{meta_learning_survey}, referred to as $D_{\text{meta,train}}$, is used to train the DNN for FSL, defined as

\begin{equation}
    D_{\text{meta,train}}=\left\{\left(D_{\text{train}}^{\text{support}},D_{\text{train}}^{\text{query}}\right)^{\left(i\right)}\right\}_{i=1}^M,  D_{\text{meta,train}}\subset D.
\end{equation}
The support set $D^{\text{support}}_{\text{train}}$ provides labeled data for the model to learn the task, while the query set $D^{\text{query}}_{\text{train}}$ measures performance on the task. Both sets consist of $N$ ways (unique classes), with $k$ shots (examples) per class. %

After meta-training, the learned \textit{meta-knowledge} is used to learn unseen tasks from a set of $Q$ tasks \cite{meta_learning_survey} provided by the target dataset $D_{\text{meta,test}}$ during meta-testing:

\begin{equation}
    D_{\text{meta,test}}=\left\{\left(D_{\text{test}}^{\text{support}},D_{\text{test}}^{\text{query}}\right)^{\left(i\right)}\right\}_{i=1}^Q, D_{\text{meta,test}}\subset D.
\end{equation}
Like before, the support set $D^{\text{support}}_{\text{test}}$ provides data to learn the new task, while $D^{\text{query}}_{\text{test}}$ is used to measure test performance on the task, with both sets potentially differing in ways and shots from meta-training.

\subsection{Evaluating Meta-Learning for On-Chip Deployment}
\label{sec:fsl_on_chip_suitability}

While a plethora of meta-learning methods exist that exploit this task-based training regime, they can be categorized into three groups: \textit{parameter initialization}, \textit{feed-forward}, and \textit{metric-learning} methods \cite{meta_learning_survey}, which we introduce with a focus on their suitability for on-chip deployment.

\begin{figure}[t]
\centering
\includegraphics[width=0.499\textwidth,trim=20 11 50 10, clip]{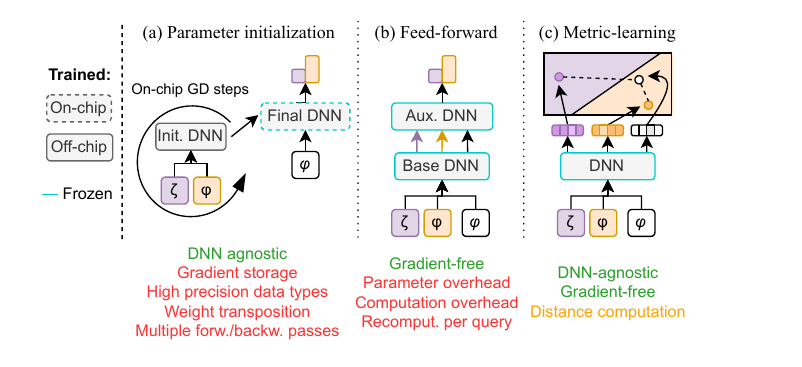}
\caption{Qualitative comparison of the suitability of the three meta-learning method types for on-chip deployment.}
\label{fig:meta_learning_comp}
\end{figure}

\textit{Parameter initialization} -- These methods aim to find optimal initial DNN parameters, enabling quick fine-tuning on unseen tasks with few examples \cite{meta_learning_survey}. Model-agnostic meta-learning (MAML) \cite{maml_finn} is a widely used approach in this category \cite{meta_learning_survey}, and employs nested GD. During meta-training, the inner GD loop is carried out for each new task, while the outer loop optimizes the initial parameters such that the inner loop requires only a few steps of GD to achieve high accuracy. Using the initial parameters resulting from meta-training, the deployment of MAML requires only the inner loop to be executed on-chip (\autoref{fig:meta_learning_comp}(a)). Key advantages are that it introduces no extra learnable parameters \cite{maml_finn} and imposes no architectural constraints on the DNN. However, a major drawback for on-chip deployment is that GD must be executed on-chip. Although proposals of initialization-based methods for edge deployment exist \cite{l2l_on_chip_marian_reptile, li2021quantization_maml, exploring_quant_in_fsl_maml}, they require (i)~buffering all intermediate activations required for GD operation, instead of discarding them as soon as the next-layer values are computed during inference, (ii)~support for on-chip weight transposition operations, and (iii)~the use of more expensive data types than quantized inference would typically require, such as (block) floating-point representations \cite{exploring_quant_in_fsl_maml,l2l_on_chip_marian_reptile}. Additionally, the proposed quantized FSL setup in \cite{l2l_on_chip_marian_reptile} requires 300 forward and backward passes to be performed for on-chip GD, significantly increasing the compute requirements for learning. Furthermore, to maintain high accuracy, \cite{l2l_on_chip_marian_reptile} resorts to the Adam optimizer \cite{kingma2014adam}, leading to a triplication of the required on-chip weight storage.

\textit{Feed-forward models} -- These methods aim to learn a direct mapping from support examples $D^{\text{support}}$ to parameters needed for predictions on query examples $D^{\text{query}}$ \cite{meta_learning_survey}. Unlike initialization-based methods, feed-forward models eliminate the need for gradient updates during deployment. However, they can introduce significant parameter overheads stemming from auxiliary DNNs or require handcrafted neural network (NN) architectures that may not align with DNNs typically supported for inference~(\autoref{fig:meta_learning_comp}(b)). A well-known example is the simple neural attentive meta-learner (SNAIL) \cite{mishra_snail_2018_abbeel}, in which the model predicts, using $N\cdot k$ support example-label pairs and a query example, the class of the query in one forward pass. SNAIL uses the same base DNN as MAML \cite{maml_finn} but requires a second, auxiliary module that has $\sim\!\!29\times$ more parameters compared to the base DNN. Another method in this category is memory-augmented NNs (MANNs) \cite{memory_augmented_neural_networks}, which use an external memory to store vector representations of previous tasks for rapid adaptation to new ones. FSL-HDnn \cite{fslhdnn} implements an on-chip MANN through hyperdimensional computing (HDC). In HDC, base DNN embeddings are projected into a high-dimensional space via multiplication with a random $\{-1,1\}$-valued matrix \cite{hdc_og_paper} (typically $>$1M parameters). Classification is then performed by finding the stored class encoding with the closest Hamming distance to the encoded input. In FSL-HDnn, a dedicated module performs encoding, storage, and classification, but increases on-chip power consumption by 120\% relative to the power for the base DNN’s forward pass.

\looseness-100 \textit{Metric learning} -- These methods aim to learn a model that generates vector representations for each support and query sample~(\autoref{fig:meta_learning_comp}(c)), which can then be used to enable the classification of unseen examples without any gradient updates, similar to feed-forward methods. A popular approach in this category is \textit{prototypical networks} (PNs) \cite{snell2017prototypical}: a feature vector (i.e.~\textit{embedding}) is computed for each sample in the support set $D^{\text{support}}$, after which these embeddings are averaged class-wise to obtain a set of \textit{prototypes}. Classification is performed by comparing the query sample embedding to the prototypes of all support classes using a \textit{distance function}%
\red{, where the L2 distance was identified as the most accurate for PNs in~\cite{snell2017prototypical}.}
The query is finally assigned to the class with the closest prototype. A key advantage of this method is that, like MAML, it is model-agnostic, requiring no additional network modules and parameters. However, unlike MAML, it is also gradient-free, with only a single extra step -- distance computation -- outside the standard NN inference pipeline. \cite{sapiens, other_fsl_cim} implement a CIM-based PN variant that stores support examples individually instead of averaging them. Only the embedding storage and distance computation/classification circuits are implemented in silicon, with embedding generation relying on off-chip floating-point DNN inference. Both designs use L1 distance, causing a 7\% accuracy drop in 32-way and 5\% in 5-way 1-shot tasks for \cite{sapiens, other_fsl_cim}, respectively, compared to an L2-based PN in FP32.

\section{The Chameleon system-on-chip architecture}
\label{sec:soc_architecture}

This section describes the Chameleon accelerator SoC, as shown in \autoref{fig:main_arch}, which highlights the key building blocks that implement our three contributions. The unified architecture for FSL, CL and inference is detailed in \autoref{sec:unified_fsl_cl_arch}, while \autoref{sec:long_context_inference_using_tcns} covers the mechanism for processing long input sequences using TCNs. Finally, \autoref{sec:matmul_free} discusses the dual-mode \red{multiplier-free} PE array.

\subsection{Unified Learning and Inference Architecture}
\label{sec:unified_fsl_cl_arch}

Chameleon’s architecture builds on the typical design of a DNN inference accelerator, featuring separate memory banks for activations (4 bits), weights (4 bits) and biases (14 bits). The activation and weight memories are connected to a $16 \times 16$ PE array tailored for efficient TCN execution.%

\begin{figure}[t]
\centering
\includegraphics[width=0.48\textwidth,trim=19 11 19 12, clip]{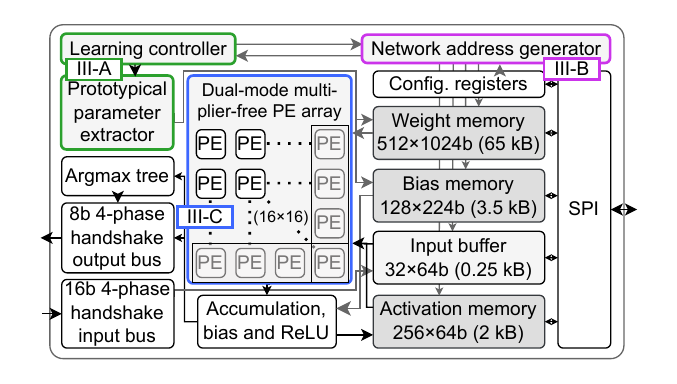}
\caption{Architecture of the Chameleon SoC, building on a typical DNN inference accelerator design. The learning controller and prototypical parameter extractor endow this architecture with on-chip learning capabilities, while the network address generator enables long-context learning and inference using TCNs. The dual-mode \red{multiplier-free} PE array allows for switching between high-throughput and low-leakage operation.}
\label{fig:main_arch}
\end{figure}

\begin{figure}[t]
\centering
\includegraphics[width=0.45\textwidth,trim=29 10 21 14, clip]{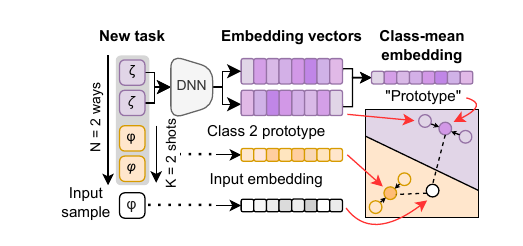}
\caption{Learning a new task using prototypes as performed in Chameleon: $N$ ways with $k$ shots are embedded using a fixed DNN to create $N\cdot k$ embedding vectors: averaging these class-wise results in $N$ prototypes. To classify an input sample, it is embedded and compared via L2 distance to the $N$ prototypes, assigning it to the class with the minimum distance.}
\label{fig:protonets_in_hardware}
\end{figure}

To endow this baseline inference accelerator with learning capabilities, we propose to leverage a reformulation of PNs (\autoref{fig:protonets_in_hardware}) as an equivalent fully-connected (FC) layer. We begin by defining each prototype as

\begin{equation}
\label{eq:prototypes}
    \bm{P}_j=\frac{\bm{s}^j}{k},~~ \bm{s}^j=\sum^k_{l=1} \bm{e}_{l,j},
\end{equation}

\noindent where $\bm{e}_{l,j}$ is the $l$th $V$-dimensional support embedding for the $j$th way and $k$ indicates the number of shots. %
Building on the insight of \cite{snell2017prototypical} by using the L2 distance function and squaring it to factor out its square root term, the PN distance computation between a prototype $\bm{P}_j$ and a query~\mbox{embedding $\bm{x}$ can be reformulated as}

\begin{equation}
\label{eq:proto_net_l2_distance_is_classification_layer}
\begin{aligned}
   \left \lVert\frac{\bm{s}^j}{k}- \mathbf{x}\right\rVert^2=D_j^2&=\sum_{i=1}^V\left(\frac{s_i^j}{k}-x_i\right)^2 \\
    &=\sum_{i=1}^V\left(\frac{{s_i^j}^2}{k^2}+\textcolor{blue}{x_i^2}-\frac{2s^j_ix_i}{k}\right).
\end{aligned}
\end{equation}

As the term $\sum_{i=1}^V\textcolor{blue}{x_i^2}$ is a constant offset that only depends on $\bm{x}$ and does not influence the \textit{relative} distances $D_j$ between $\bm{x}$ and the prototypes $\bm{P}_j$, this yields

\begin{equation}
\label{eq:middle_proto_equation}
\begin{aligned}
    D_j^2\propto  \frac{1}{k^2}\sum_{i=1}^V{s_i^j}^2-\frac{2}{k}\sum_{i=1}^Vs_i^jx_i.
\end{aligned}
\end{equation}

\noindent Then, after rescaling by $\frac{k}{2}$, we equivalently get

\begin{equation}
\label{eq:l2_is_linear}
\begin{aligned}
    D_j^2\propto \text{ }b_j+\bm{W}_j \cdot \bm{x}\text{, }~
    \text{with } b_j = \frac{1}{2k}\sum_{i=1}^V{\textcolor{red}{s_i^j}}^2&\text{,}~~\bm{W}_j = -\textcolor{red}{\bm{s}^j}.
\end{aligned}
\end{equation}

\Cref{eq:l2_is_linear} thus corresponds to an FC layer whose output neuron $j$ has a bias $b_j$ and a weight vector $W_j$, which can both be calculated from $\textcolor{red}{\bm{s}^j}$, the sum of the $k$ support embeddings as per \eqref{eq:prototypes}. \red{Although the absolute distances $D_j^2$ differ from the ones in \eqref{eq:proto_net_l2_distance_is_classification_layer}, the relative distance ordering between classes is fully preserved.}

\begin{figure}[t]
\centering
\includegraphics[width=0.48\textwidth,trim=21 11 23 12, clip]{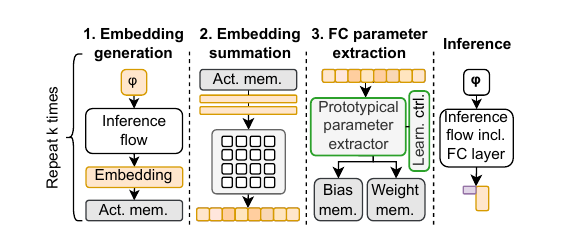}
\caption{FSL in Chameleon is performed in three steps. 1. For all $k$ shots of the new class, inference is performed to compute the embeddings, which are saved in the activation memory. 2. The embeddings from the activation memory are loaded and fed into the PE array to compute their sum. 3. This sum is converted to equivalent FC bias and weights that are stored in their respective memories. It is then possible to perform inference using the new FC layer to classify an input sample.}
\label{fig:fsl_steps_in_hardware}
\end{figure}

By reformulating PNs as an equivalent FC layer, it becomes possible to equip any inference architecture with learning capabilities with minimal overhead. To support learning, only the ability to extract the equivalent FC weight and bias parameters from a prototype is required, since the resulting FC layer can then be accelerated by the existing inference datapath. Importantly, this setup is invariant to the type of off-chip meta-training performed pre-deployment: if the final model produces high-quality embeddings, it can be deployed for on-chip FSL. Moreover, unlike initialization-based and feed-forward meta-learning methods (\autoref{sec:background}), PNs naturally extend to CL as this simply requires the storage of additional class prototypes over time. %
In Chameleon, PN weight and bias extraction is handled by two dedicated modules: the \textit{learning controller} and \textit{prototypical parameter extractor} (\autoref{fig:main_arch}, highlighted in green).
\red{The former tracks states such as the current way or shot and manages the latter, which extracts the weights and biases for new classes during learning.} 
The learning process is carried out in hardware as per \autoref{fig:fsl_steps_in_hardware}, which illustrates the three steps that Chameleon performs \red{only once} when learning each new class (way). \red{First, all shots are embedded, after which their embeddings are summed. Then, these sums are converted to FC weights and biases using the parameter extractor \red{and saved in their respective memories}: inference can now be performed for the newly learned class using these parameters.}

\red{In terms of area,} this process maximizes reuse of the inference datapath, logic, \red{and memories: as a result, it only induces a 0.5\% area overhead relative to the total core area.}
\red{The learning controller consists of only a few counters and a state machine, while the parameter extractor area scales linearly with the PE array size, independent of the number of shots, ways, or embedding dimensions. These values only affect the on-chip memory requirements, where Chameleon's small 71-kB SRAM is already sufficient to accommodate up to 250 classes.}

\red{In terms of latency, the prototypical parameter extraction} only requires $(k+2)\left\lceil\frac{V}{16}\right\rceil+1$ additional cycles \red{to learn a new class} on top of the embedding computation\footnote{~The division by 16 accounts for the $16\times16$ PE array.}, scaling linearly with both the number of shots and the embedding size. \red{For example, learning 10 new classes with 5 shots results in 0.0075\% extra latency compared to the time required to embed the shots. Similarly, inference with a new FC layer takes only $\left\lceil\frac{V}{16}\right\rceil\cdot\left\lceil\frac{N}{16}\right\rceil$ clock cycles, adding just 0.005\% latency in the 10-way scenario.}

Overall, our prototypical learning strategy enables a unified architecture for learning and inference, by reusing existing inference memories and thus eliminating the need for FSL-specific storage. It extends naturally to CL by storing new prototypes over time and can be applied to existing inference accelerator designs, with learning reduced to minimal control logic for converting prototypes into weights and biases. Unlike designs focused solely on learning, our approach enables a lightweight integration of learning into the inference process.

\subsection{Long-Context Learning and Inference Using TCNs}
\label{sec:long_context_inference_using_tcns}

Real-world natural data is often sequential, including biological signals (\textit{e.g.},~heart rate, blood pressure), environmental data (\textit{e.g.},~wind speed, humidity), and perceptual data (\textit{e.g.},~video, audio). However, the development of algorithms for FSL and CL has historically focused on benchmarks with static data, such as images~\cite{meta_learning_survey}. To the best of our knowledge, beyond a first proof of concept in~\cite{olfactory_pathway_continual_learning_cl_anp_g} for olfactory data, no chip design has demonstrated competitive performance for \textit{sequential} FSL and CL on well-established benchmarks. %

\begin{figure}[t]
\centering
\includegraphics[width=0.5\textwidth]{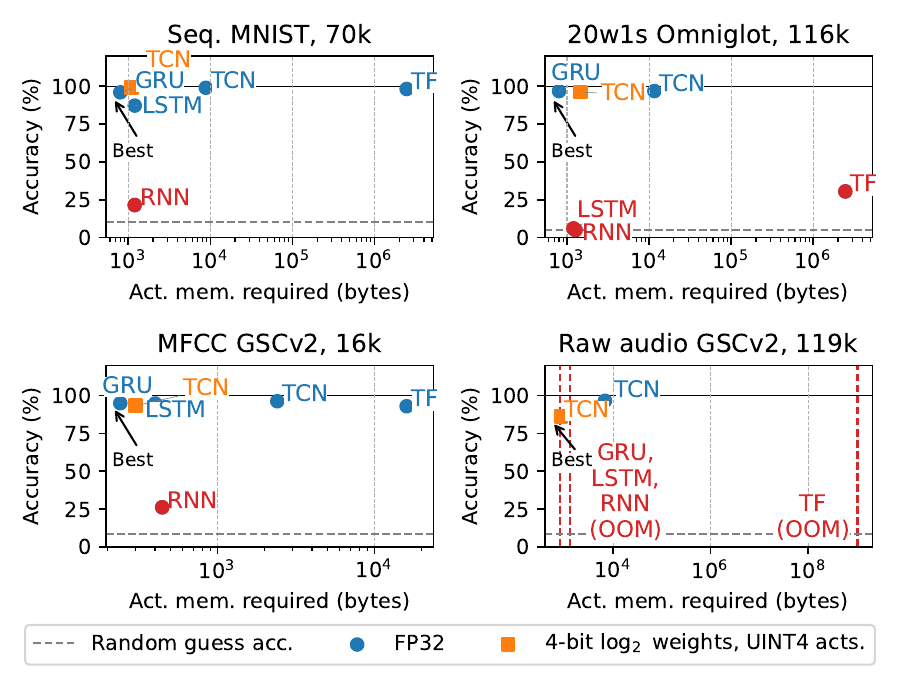}
\caption{\red{Accuracy vs. required activation memory for RNNs, LSTMs, GRUs, transformers (TFs) and TCNs at iso-parameter counts on benchmarks used for architecture selection in this work. Plain RNNs struggle with all tasks, while LSTMs and TFs fail on sequential Omniglot. GRUs closely trail TCNs in all benchmarks, but training both RNNs and TFs results in out-of-memory (OOM) errors for raw-audio KWS. Only TCNs succeed in all benchmarks in FP32, an assessment that is preserved when quantized to 4-bit $\log_2$ weights and 4-bit unsigned activations. FP32 accuracies on sequential MNIST have been taken from \cite{wang2019r}.}}
\label{fig:accuracy_vs_state_size}
\end{figure}

As our learning strategy with PNs relies on learning high-quality embeddings, we propose to extend this approach to sequential data by adopting a scalable embedder that can capture long-range dependencies across full sequences. Typical DNN models for learning with temporal data range from recurrent NNs (RNNs) to transformers \red{\cite{vaswani2017attention}}. On the one hand, RNNs have the advantage of scaling with $O(1)$ memory requirements in sequence length, at the expense of stability issues for long-range dependencies. They typically fail at $\geq1$k sequence lengths~\cite{bai2018empirical} that are necessary, \textit{e.g.},~for raw-audio processing. On the other hand, transformers~\cite{vaswani2017attention} achieve SotA performance when trained on large datasets ($>10^6$ examples), surpassing recurrent and convolutional architectures. Yet, on smaller datasets, RNNs like LSTMs~\cite{hochreiter1997long} typically outperform transformers~\cite{wang2019r}.
Furthermore, transformer encoders, required for embedding generation, also incur $O(n^2)$ memory and time complexity, making their end-to-end deployment at a \SI{}{\micro\watt}-level power budget an open challenge.
In contrast, convolutional architectures like TCNs~\cite{bai2018empirical} offer $O(\log_2 n)$ memory complexity, while iso-parameter comparisons show that they outperform both RNNs and transformers on tasks with $<10^6$ examples~\cite{wang2019r}. %
\red{In \autoref{fig:accuracy_vs_state_size}, we compare plain (Elman) RNNs \cite{elman1990finding_plain_rnn}, GRUs \cite{gru_cho}, LSTMs \cite{hochreiter1997long}, transformers (TFs) \cite{vaswani2017attention} and TCNs on relevant benchmarks for our embedder choice across sequence lengths 63, 784 and 16000. These benchmarks are sequential MNIST \cite{lecun1998gradient, le2015simple_sequential_mnist} classification, 20-way 1-shot learning on Omniglot \cite{lake_omniglot} and KWS on both MFCC features and directly on the raw data. It can be seen that in FP32, TCNs are the only architecture that succeed in all benchmarks without leading to out-of-memory (OOM) errors. They also remain competitive in both learning and inference when fully quantized, indicating the high quality of the quantized embeddings, even for long sequences.}

\begin{figure}[t]
\centering
\includegraphics[width=0.375\textwidth,trim=20 32 22 35, clip]{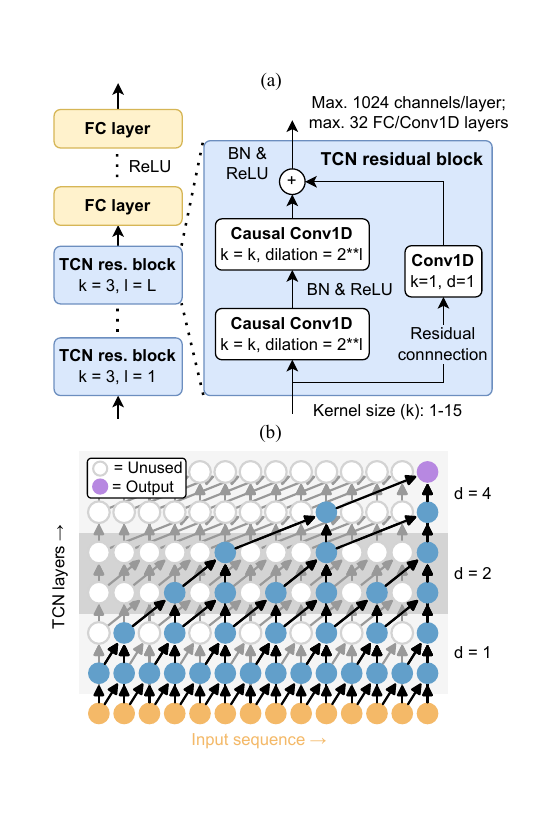}
\caption{(a)~DNN structure supported by Chameleon. Each TCN residual block contains two causal 1D convolutions, each followed by batch normalization (BN) and rectified linear unit (ReLU) activation. If input and output channels match, the Conv1D residual can be replaced with an identity. Both convolutional layers in a residual block share the same dilation factor $d$, doubling in successive blocks, starting at $d=1$. (b)~Computational graph of a 6-layer TCN (three stacked residual blocks), where white circles indicate zero-valued activations introduced by dilation.}
\label{fig:tcn_structure}
\end{figure}

Hence, we propose to use TCNs \cite{bai2018empirical} as efficient, scalable embedders for inference and FSL on sequential data at the edge. As shown in \autoref{fig:tcn_structure}(a), TCNs employ stacked causal 1D convolutions with residual connections to ease training in deep networks \cite{he2015deepresiduallearningimage} \red{while preserving the expressiveness enabled by this depth}. They are able to capture long-range dependencies via dilation, which doubles at each residual block, resulting in a receptive field ($R$) that grows \textcolor{blue}{exponentially} with network depth:

\begin{equation}
    R = 1 +\sum_{l=1}^{L/2} \textcolor{blue}{2^{l}} \cdot (k - 1),
\end{equation}
where $k$ is the kernel size and $L$ the number of layers, \red{which should be set to ensure that $R$ matches at least the input sequence length.}. In classification scenarios, dilation also makes deeper layers exponentially sparser, as illustrated by the white circles in \autoref{fig:tcn_structure}(b), resulting in a streaming memory complexity of $O(\log_2 n)$ when skipping these computations. However, leveraging this favorable memory scaling is nontrivial due to the computational graph structure induced by dilation and the residual connections in TCNs. We solve this by introducing two key techniques. %

\begin{figure*}[t]
\centering
\includegraphics[width=0.95\textwidth,trim=21 7 21 9, clip]{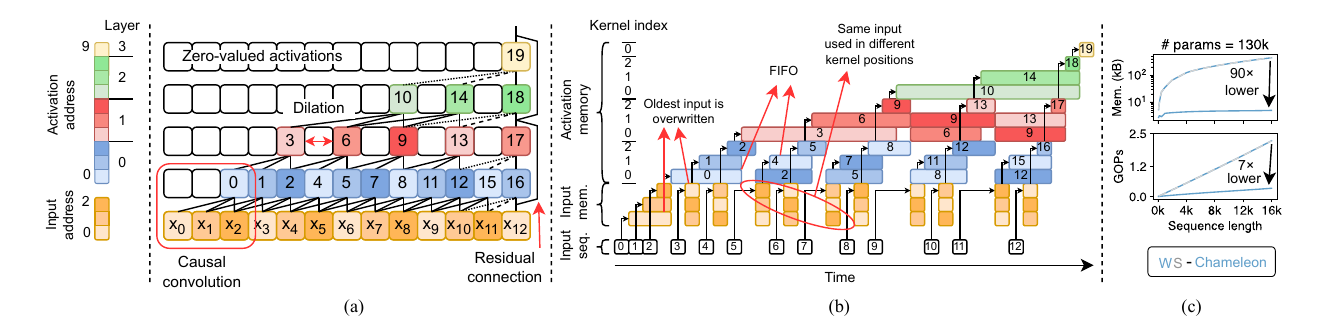}
\caption{(a)~Four-layer TCN with 13 inputs. Indices show greedy processing order; colors indicate per-layer memory locations. Causal convolutions and residuals are annotated, while the processing and storage of zero activations resulting from dilation (white) is skipped by Chameleon. (b)~FIFO-style activation memory allocation over time in Chameleon. Each layer overwrites its oldest activation during execution. Indices show processing order; bar lengths indicate activation lifetimes. (c)~Memory and compute comparison between WS TCN inference and Chameleon's strategy, using 130k parameters and producing identical outputs.}
\label{fig:tcn_struc_and_fifo_mem}
\end{figure*}

Our first technique targets the efficient exploitation of dilation-induced sparsity in the TCN's computational graph.
In TCN-CUTIE~\cite{scherer2022tcn_cutie_benini}, 1D convolutional kernels are mapped to equivalent 2D kernels: however, this introduces 80\% zero multiplications for $k=2$, as dilation is emulated via zero-padding, with sequence lengths limited to 24 timesteps for TCN processing.
UltraTrail~\cite{bernardo_ultratrail_2020} adopts a native 1D convolutional kernel mapping to perform TCN inference instead, but lacks dilation support, thereby restricting its applicability to tasks requiring small receptive fields (demonstrated up to 101 timesteps). Furthermore, its weight-stationary dataflow necessitates full sequence pre-loading, which is incompatible with streaming inference of long sequences. Giraldo~\etal~\cite{giraldo_efficient_2021} introduce a FIFO-based partial output stationary dataflow that supports dilation and reuses activations across timesteps. 
However, since the model outputs one classification per input in the sequence and as each output is connected to a different set of TCN graph nodes, this strategy does not avoid computing the unused nodes shown in~\autoref{fig:tcn_structure}(b).
Expanding on this FIFO-based approach, we introduce a \textit{greedy dilation-aware execution} (\autoref{fig:tcn_struc_and_fifo_mem}) scheme. \red{This scheme is implemented by a specialized network address generator that processes TCNs greedily through layer-wise FIFO activation storage while skipping the network's redundant, dilation-induced activations.}
\autoref{fig:tcn_struc_and_fifo_mem}(a) presents the greedy computational graph of a four-layer TCN as executed in Chameleon \red{together with the per-layer activation memory allocation}. \red{Considering the indicated computation order, it can be seen that an activation in the next layer of the TCN is only computed when the previous layer has produced all the required inputs}. When additional inputs are required, control reverts to \red{greedily} processing earlier layers, one at a time, until this requirement is met.
\red{Due to this greedy strategy, both data and structural hazards are inherently avoided.}
\red{Furthermore}, this scheme efficiently leverages our \mbox{FIFO-style} activation memory, as illustrated in \autoref{fig:tcn_struc_and_fifo_mem}(b). The address generator ensures that the output of a new timestep always overwrites the oldest, unused entry in that layer's FIFO buffer. The same mechanism applies to the input memory, where outdated inputs are automatically replaced. \red{Since the computational graph of the deployed TCN is known in advance, operations are scheduled following a predefined layer-wise schedule to prevent FIFO underflows.}
\red{This greedy processing is then combined with dilation-aware operation, which skips computations for zero-valued activations introduced by dilation that do not contribute to the final output.} \red{This combination of techniques enables the memory required for TCN inference to scale linearly with depth. However, since the TCN's receptive field doubles with each additional layer, the total activation storage increases only with $\log_2$ of the receptive field size.}
\red{These scaling properties directly translate into practical gains}: we demonstrate in~\autoref{fig:tcn_struc_and_fifo_mem}(c) that Chameleon achieves $90\times$ memory and $7\times$ compute reduction over weight stationary (WS) non-dilation-optimized TCN inference at a sequence length of 16k (raw audio) with a fixed network size of 130k parameters (Chameleon’s maximum supported size).

\begin{figure}[t]
\centering
\includegraphics[width=0.48\textwidth,trim=80 15 30 16, clip]{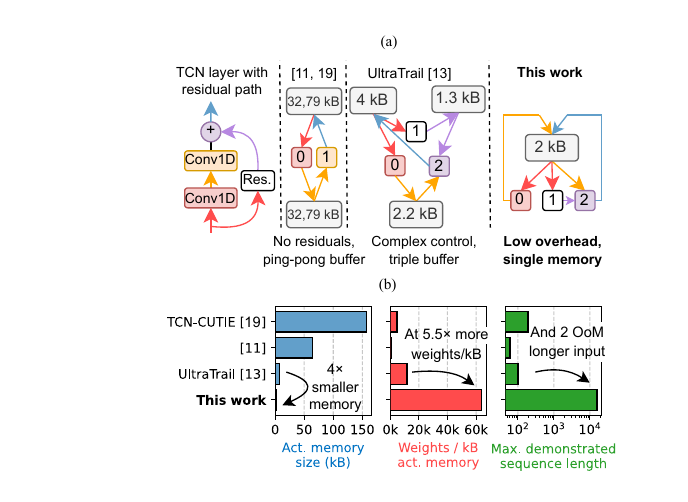}
\caption{(a)~Comparison of residual operation steps and activation memory sizes across TCN accelerators. \cite{giraldo_efficient_2021,scherer2022tcn_cutie_benini} use ping-pong buffers but lack residual layer support; \cite{bernardo_ultratrail_2020} requires a triple buffer setup for residual layers; Chameleon uses only a single dual-port memory, reducing memory and control complexity. (b)~Impact of computation strategy on activation memory size, maximum number of weights/\SI{}{\kilo\byte} of activation memory and input sequence length between TCN accelerators.}
\label{fig:tcn_processing_comparison}
\end{figure}

\looseness-100 Our second technique enables efficient handling of residual connections. As sequence lengths grow, larger dilation factors stemming from deeper networks are needed to expand the receptive field (\autoref{fig:tcn_structure}), making residual connections essential for effective optimization of these deep networks \cite{he2015deepresiduallearningimage}. While TCN-CUTIE \cite{scherer2022tcn_cutie_benini} supports up to ten convolutional layers, it lacks support for residual connections, severely limiting scalability. UltraTrail \cite{bernardo_ultratrail_2020} does support residuals but requires three separate buffers and complex memory access control logic. %
Giraldo~\etal~\cite{giraldo_efficient_2021} omit residuals but still employ a ping-pong buffer setup. To eliminate the need for multi-buffer schemes to manage residual layers without sacrificing FIFO-style inference, our \textit{residual-aware register file processing} (\autoref{fig:tcn_processing_comparison}(a), right) uses a two-port register file that enables simultaneous reads and writes. \red{When the next layer needs the previous layer’s output in the same cycle the latter computed, the network address generator inserts a one-cycle delay to avoid read/write conflicts. This negligible 1-cycle latency overhead would only happen for a tiny $16\times 16$ FC layer immediately followed by another FC layer, and does not occur for practical network sizes such as the ones used in this paper. Finally, in order} to support asynchronous streaming input alongside ongoing processing, Chameleon stores the incoming DNN inputs in a dedicated \SI{0.25}{\kilo\byte} memory.

By combining the two techniques described above and as shown in \autoref{fig:tcn_processing_comparison}(b), Chameleon reduces activation memory by $76\times$, $28\times$, and $4\times$ compared to prior TCN accelerators \cite{scherer2022tcn_cutie_benini, giraldo_efficient_2021, bernardo_ultratrail_2020}, while enabling deployment of DNNs with up to $5.5\times$ more weights per \SI{}{\kilo\byte} of activation memory. This enables, for the first time, raw audio KWS on 16k-step inputs --- two orders of magnitude longer than in previous TCN accelerators.

\subsection{High-Throughput or Low-Leakage Operation Using a Dual-Mode, \red{Multiplier-Free} PE Array}
\label{sec:matmul_free}

Since learning in Chameleon is integrated within the inference pipeline, we propose two architectural enhancements, so as to (i) reduce the PE array and memory footprints with inference-optimized quantization, allowing for \red{multiplier-free} operation with 4-bit signed $\log_2$ weights, and (ii) endow the PE array with the flexibility to optimally support different inference and learning scenarios. 

\begin{figure}[t]
\centering
\includegraphics[width=0.49\textwidth,trim=23 20 20 19, clip]{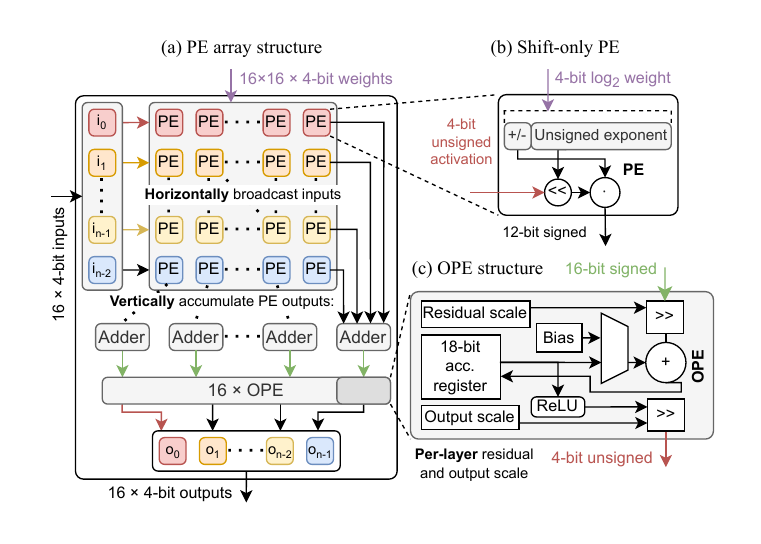}
\caption{(a)~Diagram of the \red{multiplier-free} PE array, operating under an output-stationary dataflow. (b)~Individual PE, which uses a bit shift in place of a multiplication. (c)~Output PE (OPE) module, which performs input/output rescaling as well as bias addition and ReLU activation.
}
\label{fig:wide_pe_array_figure}
\end{figure}

\begin{table}[t]
\caption{\red{Quantization accuracy assessment on 20-way Omniglot tasks using 1, 5 and 10 shots with a 116k-parameter TCN. For all quantized experiments, we apply the FC layer conversion of the prototypes per \eqref{eq:l2_is_linear}. The ± shows 95\% confidence intervals over 100 tasks, while the bold font indicates the best accuracy when fully quantized.}}
\label{tab:int4_int8_log4}
\centering
\newcommand{\thline}{\specialrule{1pt}{0pt}{0pt}}
\begin{adjustbox}{max width=0.48\textwidth}
\begin{tabular}{@{}cccccc@{}}
\thline
\multicolumn{3}{c}{Quantization format} & \multicolumn{3}{c}{20-way accuracy} \\ \thline
Weight & Act. & \multicolumn{1}{c|}{Proto.} & 1-shot & 5-shot & 10-shot \\ \hline
FP32 & FP32 & \multicolumn{1}{c|}{FP32} & 94.5 ± 0.5\% & 98.4 ± 0.2\% & 98.7 ± 0.2\% \\
INT8 & FP32 & \multicolumn{1}{c|}{FP32} & 96.1 ± 0.4\% & 98.8 ± 0.3\% & 98.9 ± 0.3\% \\
INT8 & UINT4 & \multicolumn{1}{c|}{FP32} & 88.9 ± 0.8\% & 97.0 ± 0.3\% & 97.8 ± 0.3\% \\
INT8 & UINT4 & \multicolumn{1}{c|}{UINT7} & 86.4 ± 0.9\% & 94.8 ± 0.4\% & 96.2 ± 0.3\% \\
INT4 & UINT4 & \multicolumn{1}{c|}{UINT7} & \textbf{89.2 ± 0.8}\% & 95.8 ± 0.4\% &\textbf{ 96.9 ± 0.3\%} \\
INT4 & UINT4 & \multicolumn{1}{c|}{UINT3} & 86.8 ± 0.8\% & 91.9 ± 0.6\% & 93.0 ± 0.6\% \\
$\log_2$-4 & UINT4 & \multicolumn{1}{c|}{UINT3} & 89.0 ± 0.8\% & 94.2 ± 0.5\% & 95.1 ± 0.4\% \\
$\log_2$-4 & UINT4 & \multicolumn{1}{c|}{$\log_2$-4} & \textbf{89.1 ± 1.3}\% & \textbf{96.1 ± 0.5\%} & \textbf{96.8 ± 0.3\%} \\ \thline
\end{tabular}
\end{adjustbox}
\end{table}

Our first technique focuses on \textit{inference-optimized quantization}. For instance, proposals for deploying parameter initialization meta-learning methods on-chip require GD, which entails floating-point precision requirements. Murthy~\etal~\cite{l2l_on_chip_marian_reptile} use block floating point (8-bit shared exponent, 4-bit mantissa) for weights, activations, gradients and errors. Compared to regular integer operation, additional logic for exponent management, as well as normalization and alignment is required. Wang~\etal~\cite{exploring_quant_in_fsl_maml} demonstrates the feasibility of using signed 16-bit uniform quantization of the weights and gradients for various initialization-based meta-learning methods, but reduces memory requirements only by $2\times$ compared to learning with FP32. FSL-HDnn \cite{fslhdnn} uses brain floating point (Bfloat)-16 to generate embeddings for HDC during FSL, also saving $2\times$ memory compared to a vanilla FP32 implementation. A deployment proposal for MAML~\cite{li2021quantization_maml} demonstrates the use of 4-bit signed uniform weights, but still requires 8- or 16-bit gradients depending on task difficulty, hence increasing memory requirements by $2-4\times$ compared to inference-only quantization.

\begin{figure}[t]
\centering
\includegraphics[width=0.37\textwidth,trim=21 23 21 23, clip]{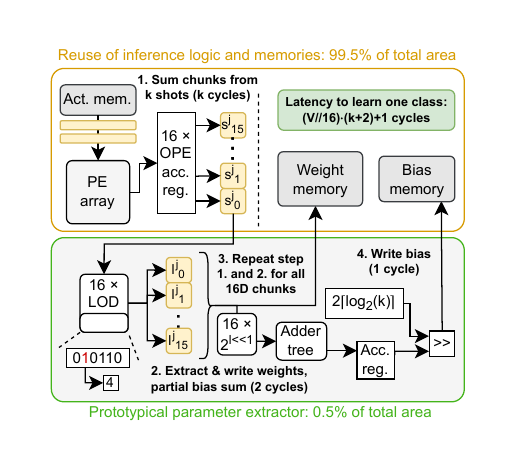} \caption{\red{Detailed circuit-level view of the multiplier-free prototypical parameter extractor, which includes a leading-one detector (LOD) for the $\lfloor\log_2\rfloor$ operation. Additionally, the extensive reuse of the inference hardware for learning is illustrated. Per extraction operation, the latency in clock cycles is annotated.}} \label{fig:proto_param_extractor} \end{figure}

As Chameleon operates in a purely forward manner due to its gradient-free learning, it can rely on a simple quantization scheme that is optimized for inference. Hence, we propose to use $\log_2$ weights \cite{miyashita2016convolutional_log2_pot2_power_of_two}: this allows replacing \red{the} multiplier \red{in each PE} with a bit shifter, \red{thereby enabling} \red{multiplier-free} inference and embedding computation (\autoref{fig:wide_pe_array_figure}(a)).
\red{We show the FSL performance for various quantization scenarios in \autoref{tab:int4_int8_log4}, using the Omniglot dataset \cite{lake_omniglot}, where the
activation quantization impacts accuracy much more than weight quantization. Yet, significant savings can be achieved: the 4-bit $\log_2$ ($\log_2$-4) quantization scheme limits the accuracy penalty compared to the FP32 baseline to only 1.9--2.3\% in the 10- and 5-shot cases, and outperforms INT8 and INT4 weight quantization. 
The 1-shot case is more sensitive to quantization due to reduced prototype precision: while this is illustrated by an increased accuracy degradation against FP32 at 5.4\%, performance still compares favorably to state-of-the-art chip implementations, as we will show in \autoref{sec:fsl_and_cl_measurements}.}

\red{Replacing the 4-bit multiplier inside each PE with a shifter not only improves accuracy, but also reduces PE area by 42\%. The} $\log_2$ PEs (\autoref{fig:wide_pe_array_figure}(b)) first left-shift the horizontally-broadcast input by the weight's exponent and then apply a sign correction. These outputs are summed vertically and accumulated in 18-bit signed uniform registers within output PEs (OPEs), detailed in \autoref{fig:wide_pe_array_figure}(c). The OPE performs input rescaling (for residuals), bias addition, ReLU activation, and output rescaling before its output is written back to the activation memory. Finally, the PE array adopts an output-stationary dataflow during inference, to efficiently support our greedy TCN processing scheme. Each cycle, it receives 16 4-bit unsigned uniform ReLU-based activations, either from the input memory (first layer) or activation memory (subsequent layers), along with $16\times16$ $\log_2$-quantized weights from the weight memory.

With inference being fully \red{multiplication-free, we aim to remove the remaining multiplications during learning by also adapting} Eq.~\eqref{eq:l2_is_linear} to a $\log_2$ formulation. \red{This allows for efficient PN parameter extraction with improved FSL accuracy: \autoref{tab:int4_int8_log4} shows that $\log_2$-4 prototype quantization unilaterally outperforms UINT, while ensuring that the few-shot-learned FC layer runs on existing bit-shift PEs, avoiding a dedicated $16\times16$ multiplier array.}
\red{To this end, each support embedding sum} ($\textcolor{red}{\bm{s}^j}$) is first $\log_2$-quantized, \red{which removes the square in the bias term of Eq.}~\eqref{eq:l2_is_linear}. \red{As such, the remaining multiplications in the datapath can be replaced by bit shifts:}

\begin{equation}
\label{eq:l2_is_linearshifted}
\begin{aligned}
    \textcolor{red}{\bm{l}^j} = \left\lfloor\log_2\textcolor{red}{\bm{s}^j}\right\rfloor,~~
    b_j = \frac{1}{2k}\sum_{i=1}^V{2^{\textcolor{red}{l_i^j}<<1}}&\text{,}~~\bm{W}_j = 2^{\textcolor{red}{\bm{l}^j}}.
\end{aligned}
\end{equation}

\noindent \red{All operations in the above equation, consisting only of bit shifts and additions, are implemented on-chip as part of the prototypical parameter extractor: \autoref{fig:proto_param_extractor} shows its detailed circuit diagram.} Note that the division by $2k$ in Eq.~\eqref{eq:l2_is_linearshifted} is approximated by a right shift with $2\left\lceil \log_2(k) \right\rceil$ bits \red{while} $\left\lfloor\log_2\textcolor{red}{\bm{s}^j}\right\rfloor$ \red{is computed by extracting the most significant bit's position using a leading-one detector (LOD).}

Our second technique allows \textit{efficiently supporting different deployment scenarios}, from leakage-dominated real-time KWS inference to throughput- and dynamic-power-constrained FSL operation with large DNN embedders. For instance, Vocell~\cite{vocell} has a 36\% leakage contribution (excluding analog feature extraction) during real-time KWS, similar to the 30\% for UltraTrail \cite{bernardo_ultratrail_2020} and $\sim$33\% for Giraldo~\etal~\cite{giraldo_efficient_2021} with limited throughputs of 0.13, 3.8 and 0.26 GOPS, respectively. 
In contrast, TinyVers~\cite{vikram_tinyvers} achieves a $4-135\times$ higher throughput of 17.6 GOPS. However, this comes at the cost of an order-of-magnitude increase in real-time KWS power consumption.
Hence, to support low-leakage operation without degrading the design's throughput, we introduce a dual-mode PE array whose size is reconfigurable, which enables Chameleon to efficiently support \textit{both} \SI{}{\micro\watt}-power real-time inference and high-throughput tasks with large DNN embedders.

\begin{figure}[t]
\centering
\includegraphics[width=0.46\textwidth,trim=20 17 40 20, clip]{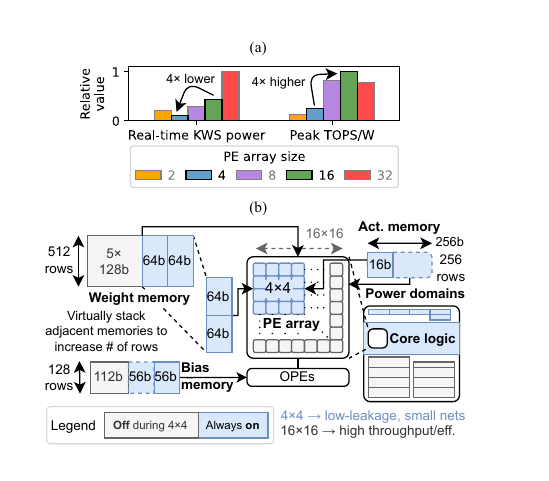}
\caption{(a)~Comparison of simulated real-time KWS power and peak TOPS/W estimates for different PE array sizes. (b)~Data layout for activation, bias, and weight memories to support both $4\times4$ and $16\times16$ PE array usage, allocating LSBs of weight memories to the top-left $4\times4$ section. In $4\times4$ mode, gray memories are powered off, while the remaining weight memories are virtually stacked to double the row count.}
\label{fig:subsection_mode}
\end{figure}

\begin{figure}[t]
\centering
\includegraphics[width=0.499\textwidth,trim=21 10 21 10, clip]{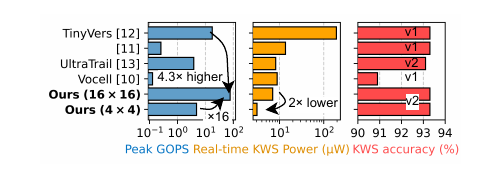}
\caption{Peak GOPS, real-time KWS power and accuracy comparison on 12-class Google Speech Commands (GSC, v1 and v2) between KWS accelerators.}
\label{fig:tcn_accel_comp}
\end{figure}

To determine the optimal PE array sizes for the two modes, we simulate real-time KWS \red{power} and peak TOPS/W performance \red{during FSL} in \autoref{fig:subsection_mode}(a), assuming SRAMs dominate total system power. The analysis identifies array sizes of 4 and 16 as optimal. \red{To support efficient operation for both sizes, we propose to use a single $16\times16$ PE array while we employ a special memory layout: instead of storing DNN weights purely in row- or column-major order, the top-left $4\times4$ weights are stored row-major in the first two banks (LSBs), followed by the remaining \red{240} weights (\autoref{fig:subsection_mode}(b)). This approach has two key advantages. First, by writing the network parameters for the $4\times4$ mode only to the LSB memories, referred to as \textit{always-on} banks, the uninitialized MSB memories can be power-gated through a separate power rail that is physically powered off, thereby cutting leakage without data loss. The 240 unused PEs can also be clock gated to reduce dynamic power.} \red{Second, the design of the network address generator can remain mode-independent: in both modes, the addresses are generated in the same way; the only difference is that in the $4\times4$ mode, a smaller weight matrix is loaded per cycle. Hence, as each address now contains fewer parameters, we virtually stack the LSB weight banks in $4\times4$ mode to enlarge the address space, supporting 16k weights over 1024 addresses, instead of 512 in $16\times16$ mode.}

Comparing Chameleon to other KWS accelerators in \autoref{fig:tcn_accel_comp}, we observe that its dual-mode operation enables a $4.3\times$ higher peak throughput in $16\times16$ mode, a 16-fold increase from Chameleon's $4\times4$ mode. Meanwhile, in $4\times4$ mode, Chameleon achieves a $2\times$ power reduction for real-time end-to-end KWS inference while maintaining a classification accuracy on par with state-of-the-art inference-only accelerators. \red{Varying the PE array size only impacts the throughput-power tradeoff, without impacting accuracy}.

\section{Measurements}
\label{sec:measurements}

Chameleon was taped out in TSMC 40-nm low-power (LP) CMOS technology%
~(\autoref{fig:annotated_die_photograph}). It occupies \SI{1.25}{\mm\squared} with a core area of \SI{0.83}{\mm\squared}, including the power rings. The SoC has two separate power domains, corresponding to (i) the core logic (using a mix of SVT and HVT standard cells) and always-on memories, and (ii) the gateable MSB memories.

We first outline the software and measurement setup (\autoref{sec:test_setup}), then present the results for FSL and CL (\autoref{sec:fsl_and_cl_measurements}), followed by real-time KWS results (\autoref{sec:real_time_kws}), and finally compare Chameleon to the SotA (\autoref{sec:comp_with_sota}).

\subsection{Software and Measurement Setup}
\label{sec:test_setup}

Our TCN models are based on the original architecture in~\cite{bai2018empirical}, where we introduce several modifications incorporating techniques from modern deep convolutional residual networks (ResNets)~\cite{he2015deepresiduallearningimage} for normalization~\cite{ioffe2015batchnormalizationacceleratingdeep,goyal2017accurate} and initialization~\cite{goyal2017accurate,he2015delving}.

To deploy TCNs on Chameleon, we perform quantization-aware training (QAT) using the Brevitas framework \cite{brevitas}, after FP32 training. QAT starts from the FP32 checkpoint with the lowest validation loss, with all parameters quantized and batch normalization (BN) layers folded into the weights of the previous layer, as per \cite{batch_norm_folding_krishnamoorthi2018quantizing, batch_norm_folding_Jacob_2018_CVPR}. Within Brevitas, we implemented a custom variable-bit quantizer for signed $\log_2$ weights and a custom requantizer to simulate overflow of the 4-bit unsigned uniform activations in Chameleon. %
We apply asymmetric and symmetric quantization for activations and weights respectively, both with per-tensor scaling \cite{nagel2021whitepaperneuralnetwork}. During quantized PN training, the TCN embeddings are using 4-bit unsigned uniform quantization, like activations (see \autoref{sec:matmul_free}). The resulting prototypes are quantized using 4-bit signed $\log_2$ quantization, as they are converted to prototypical parameters as per Equation~\eqref{eq:l2_is_linearshifted}.

\begin{figure}[t]
\centering
\includegraphics[width=0.499\textwidth,trim=30 30 15 24, clip]{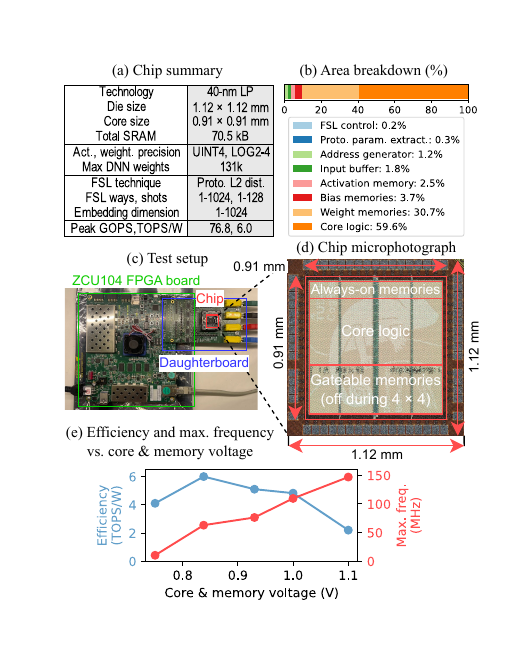}
\caption{(a)~Metrics summary of the Chameleon SoC. (b)~Area breakdown of the SoC per module. (c)~Test setup, including a Zynq UltraScale MPSoC ZCU104 FPGA that connects to a daughterboard hosting the chip. (d)~Annotated die photograph of the Chameleon SoC. The bottom section of the SoC, containing the gateable memories, can be powered off to enable inference with small networks using the $4\times4$ mode of the PE array. (e) Efficiency and maximum frequency characterization of Chameleon.}
\label{fig:annotated_die_photograph}
\end{figure}

Our measurement setup is shown in~\autoref{fig:annotated_die_photograph}(c). A daughter board hosting the Chameleon chip connects to external power supplies and is interfaced with a Zynq UltraScale MPSoC ZCU104 evaluation board, which provides test stimuli and records the results. The complete deployment and test setups are part of the open-source repository of Chameleon. %
Note that all metrics that we report in the remainder of this section are obtained at room temperature.

\subsection{FSL and CL}
\label{sec:fsl_and_cl_measurements}

To demonstrate end-to-end on-chip learning on Chameleon using our unified learning and inference architecture, we benchmark Chameleon for FSL and CL scenarios.

\begin{figure}[t]
\centering
\includegraphics[width=0.5\textwidth]{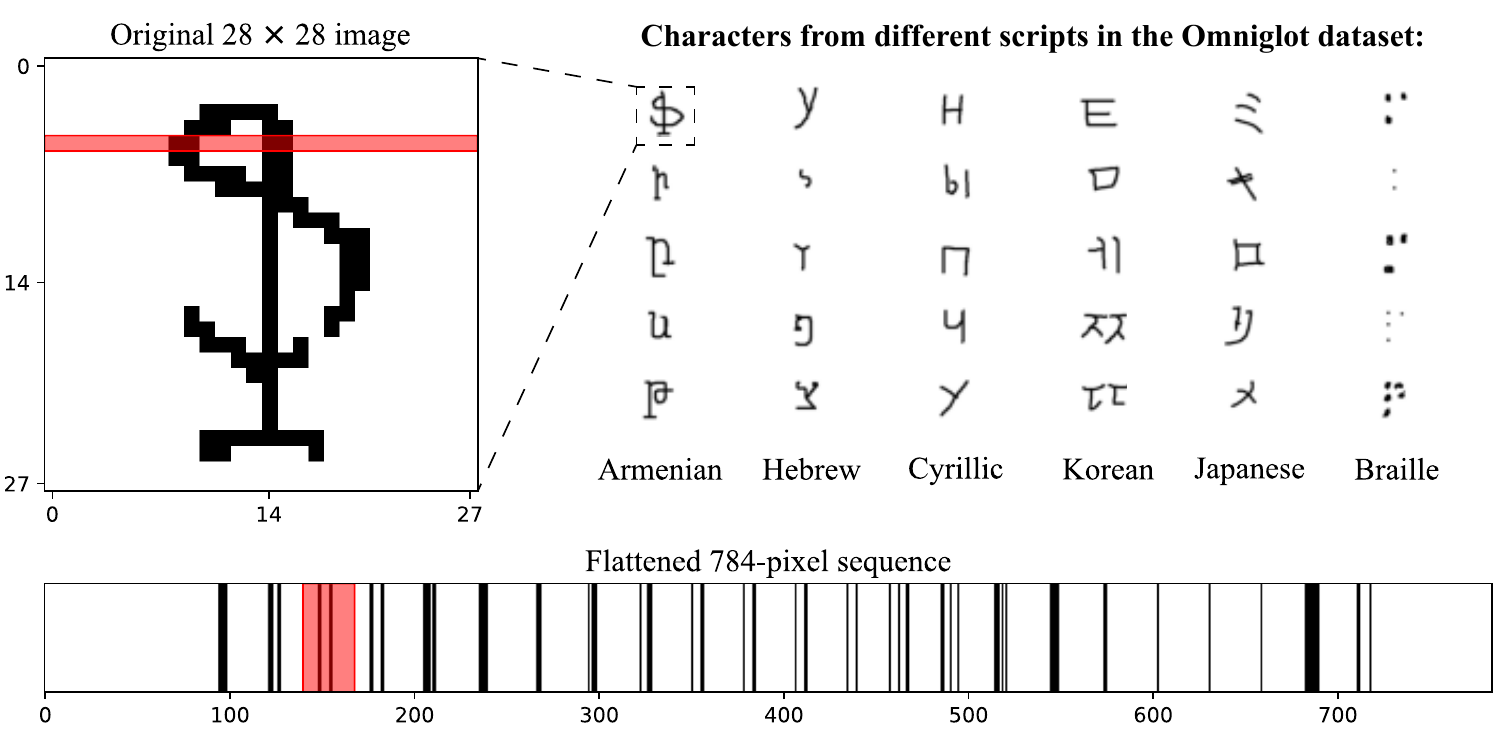}
\caption{Sample characters from different alphabets, taken from the Omniglot dataset, including a demonstration of flattening an image from the dataset to enable \textit{sequential Omniglot} on 1D sequences of pixel data.}
\label{fig:sequential_omniglot}
\end{figure}

\textit{Benchmarking setup} -- For both FSL and CL, we use the Omniglot dataset \cite{lake_omniglot}, which comprises 1,623 classes of handwritten characters from 50 alphabets (see~\autoref{fig:sequential_omniglot}, right, for selected examples). Every class has 20 examples, each of which is drawn by a different person. To align with other works on FSL, we use the Vinyals train-val-test split \cite{vinyals2016matching}. %
Following \cite{memory_augmented_neural_networks, snell2017prototypical}, we augment all splits by generating new classes via 90\degree, 180\degree, and 270\degree \;rotations and resize all images to $28\times28$ for consistency. To adapt the images for processing with TCNs, we flatten each image pixelwise (see~\autoref{fig:sequential_omniglot}, bottom), effectively creating a \textit{sequential Omniglot} representation. \red{While this dataset lacks the temporal complexity of real-world sequential data, it has the benefit of offering a reliable point of comparison with previous work, which we will later complement with the use of the Google Speech Commands Dataset in \autoref{sec:real_time_kws}.}

\textit{FSL results} -- Using a 116k-parameter TCN with 14 layers, we report in \autoref{tab:smal_fsl_table} the classification accuracies on the Omniglot test set across 5,20-way 1,5-shot as well as 32-way 1-shot scenarios found in the state of the art. It can be seen that Chameleon outperforms existing works \cite{other_fsl_cim, sapiens, fslhdnn} by up to 16 accuracy points, setting new records across all scenarios, even though both FSL CIM designs \cite{other_fsl_cim, sapiens} use FP32 off-chip embedders.
Chameleon consumes \SI{11.6}{\milli\watt} for end-to-end FSL on Omniglot at \SI{100}{\mega\hertz} and \SI{1.0}{\volt}. At \SI{100}{\kilo\hertz} and \SI{0.625}{\volt}, it consumes \SI{12.9}{\micro\watt}. The latencies for learning one shot of a new class are \SI{0.59}{\milli\second} and \SI{0.54}{\second} respectively, yielding an energy per shot of \SI{6.84}{\micro\joule} and \SI{6.97}{\micro\joule}, including the embedding phase.
Since prototypical parameter extraction only takes a few clock cycles, $<\!\!0.02\%$ of the embedding computation time on Omniglot, learning and inference have effectively identical total energy, power and latency. This contrasts sharply with other on-chip training methods like CHIMERA~\cite{chimera_backprop}, which employs low-rank training with 8-bit signed parameters and requires $10^3$–$10^4$ RRAM update steps for learning. Similarly, Park~\etal~\cite{park_backprop_float} use FP8 for full on-chip SGD-based training and hence could host MAML-like setups. However, their lowest reported power is already 13\% higher than Chameleon's peak end-to-end power, demonstrating its incompatibility with extreme-edge operation. Alternatively, instead of learning new classes, Cioflan~\etal~\cite{cioflan2024boostingkeywordspottingondevice} use embeddings %
to adapt to the speech characteristics of end users for KWS. A microcontroller updates an embedding layer using few-shot spoken keywords, inducing a 35\% FLOPS overhead compared to embedding computation alone. %

\begin{table}[t]
\caption{FSL test accuracy comparison between FSL accelerators that have reported results in silicon on the Omniglot dataset. The ± shows 95\%
confidence intervals over 100 tasks.}
\label{tab:smal_fsl_table}
\centering
\begin{adjustbox}{max width=0.48\textwidth}
\begin{threeparttable}
\addtolength{\tabcolsep}{-0.4em} %

\begin{tabular}{@{}rccccc@{}}
\toprule
\multicolumn{1}{l}{} & \multicolumn{2}{c}{5-way} & \multicolumn{2}{c}{20-way} & 32-way \\
\multicolumn{1}{l}{} & 1-shot & 5-shot & 1-shot & 5-shot & 1-shot \\ \midrule
Kim~\etal$^*$~\cite{other_fsl_cim} & 93.4\% & 98.3\% & - & - & - \\
SAPIENS$^*$ \cite{sapiens} & - & - & - & - & 72\% \\
FSL-HDnn \cite{fslhdnn} & 79.0\% & - & - & 79.5\% & - \\
\textbf{This work} & \textbf{\omnfwayoshot} & \textbf{\omnfwayfshot} & \textbf{\omntwayoshot} & \textbf{\omntwayfshot} & \textbf{\omnthwayoshot} \\ \bottomrule
\end{tabular}
\begin{tablenotes}
  \item[*] Uses an off-chip FP32 embedder.
  \end{tablenotes}
\end{threeparttable}

\end{adjustbox}
\end{table}

\begin{figure}[t]
\centering
\includegraphics[width=0.49\textwidth,trim=0 11 0 0, clip]{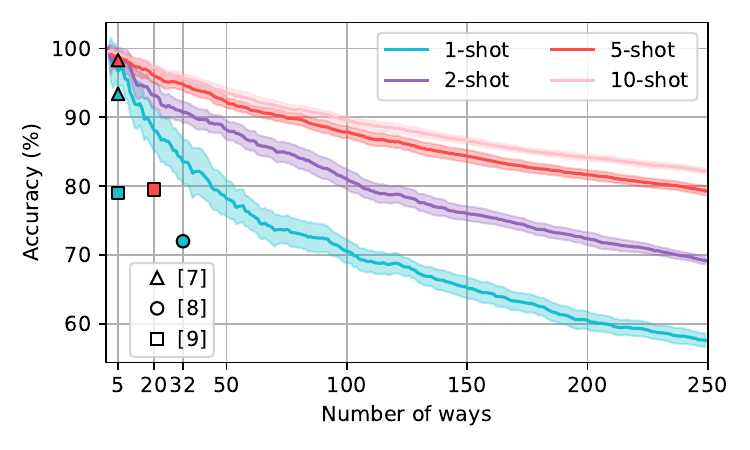}
\caption{\red{Average} end-to-end CL classification accuracies on the Omniglot dataset using Chameleon for 2\textendash250 ways with 1, 2, 5 and 10 shots, compared to other FSL chips (points outlined in black). The shaded regions indicate 95\% confidence intervals over 20 tasks. For $<5$ ways, shot count has little effect on accuracy; with more ways, additional shots help, \red{with diminishing returns} beyond 5 shots.}
\label{fig:cl_on_ominglot}
\end{figure}

\begin{table}[t]
\caption{\red{Accuracy (ACC) and BWT metrics for continually learning 250 classes using 1, 2, 5 or 10 shots. The metrics are averages over 20 tasks. The ± indicates 95\% confidence intervals over these tasks.}}
\label{tab:fwt_bwt}
\centering
\begin{tabular}{@{}rcccc@{}}
\toprule
 & 1-shot & 2-shot & 5-shot & 10-shot \\ \midrule
ACC & 57.6 ± 0.8\% & 68.9 ± 0.6\% & 79.1 ± 0.5\% & 82.2 ± 0.5\% \\
BWT & -0.127 & -0.104 & -0.076 & -0.067 \\ \bottomrule
\end{tabular}
\end{table}

\textit{CL results} -- We evaluate CL on Chameleon using the same TCN model as for FSL. CL is performed by learning one new class at a time, up to 250 classes, using 1, 2, 5, or 10 shots. As Chameleon is the only silicon-proven work to perform end-to-end fully on-chip CL, \autoref{fig:cl_on_ominglot} compares its CL performance (continuous lines) to the FSL performance of prior works~\cite{other_fsl_cim, sapiens, fslhdnn} (single data points). %
\red{In \autoref{tab:fwt_bwt}, we also report the average accuracy (ACC, equivalent to final accuracy in \autoref{fig:cl_on_ominglot}) and backward transfer (BWT) metrics \cite{lopez2017gradient} for continually learning 250 classes using Chameleon.
We omit the forward transfer (FWT) metric, as it is not informative because the PNs implemented in Chameleon do not support zero-shot learning \cite{lopez2017gradient}.
The slightly negative BWT values indicate that learning a new task decreases the performance on a preceding task, indicating forgetting, which is consistent with the curves in \autoref{fig:cl_on_ominglot}.%
}
Notably, in the similar few-shot class-incremental learning (FSCIL) \cite{tao2020few_fscil_og_authors_x} scenario, where a DNN learns a set of classes before deployment and a new set of classes online, there is currently also no scalable end-to-end fully on-chip design.
For example, Karunaratne~\etal~\cite{karunaratne2022memory_continual_learning_hermes_cl} propose a CIM macro for MANNs targeting FSCIL: their design supports embedding aggregation (summing) and uses a $\sim\!4\times$ larger embedder than \cite{other_fsl_cim}, enabling 100-way FSCIL with 5 shots per class. However, embeddings are still generated off-chip with an FP32 embedder and binarized before on-chip use, similar to~\cite{other_fsl_cim, sapiens}.
Alternatively, Wibowo~\etal~\cite{wibowo202412_microcontroller_cl_continual_learning_cioflan_benini} demonstrate FSCIL with the embeddings for PNs computed on-chip, using a microcontroller. However, they require an external L3 memory to store the 8-bit ($2\times$ larger than Chameleon) DNN weights and use cosine similarity as the distance function for PNs, requiring an expensive division operation.
\red{More recently, Song~\etal~proposed CLO-HDnn \cite{clo_hdnn}, building on the FSL-HDnn accelerator \cite{fslhdnn}, to target CL using HDC. To reduce computations, the authors introduce a bypass mechanism to only activate their embedder network for difficult tasks. However, similar to FSL-HDnn \cite{fslhdnn}, the design does not support fully end-to-end on-chip operation: only 0.14\% of the required DNN weights can be stored on-chip at any time. Furthermore, CLO-HDnn only supports up to 128 classes, compared to 250 for Chameleon, even though the design's total on-chip memory and total silicon area are $\sim3\times$ and $\sim20\times$ larger than Chameleon respectively.}
Beyond embedding-based learning, in~\cite{olfactory_pathway_continual_learning_cl_anp_g}, Huo~\etal~demonstrate a first proof of concept for fully on-chip end-to-end FSCIL. The design uses a spiking NN with spike timing-dependent plasticity and lateral inhibition: it supports classification of nine types of gas using olfactory data and overcomes data disturbances caused by sensor drift. However, since it supports networks of only up to 4k 4-bit parameters, its suitability for more complex tasks such as Omniglot is unclear.

\begin{figure}[t]
\centering
\includegraphics[width=0.5\textwidth,trim=0 6 0 7, clip]{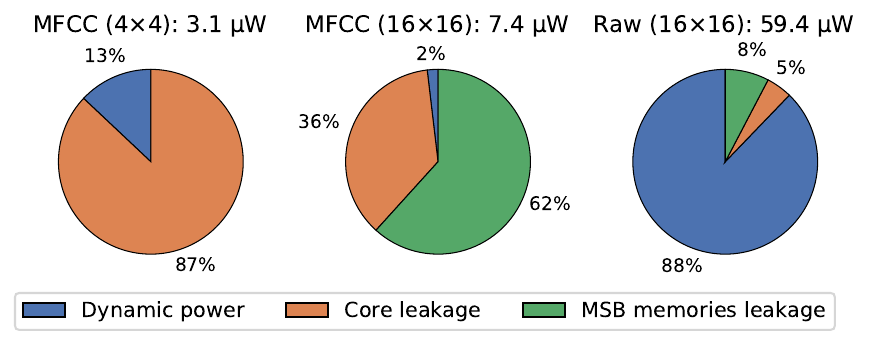}
\caption{Comparison of power contributions (at \SI{0.73}{\volt}) during real-time KWS using MFCC feature vectors in both $4\times4$ and $16\times16$ mode, as well as raw audio in $16\times16$ mode.}
\label{fig:power_comparison_kws}
\end{figure}

\begin{figure}[t]
\centering
\includegraphics[width=0.49\textwidth]{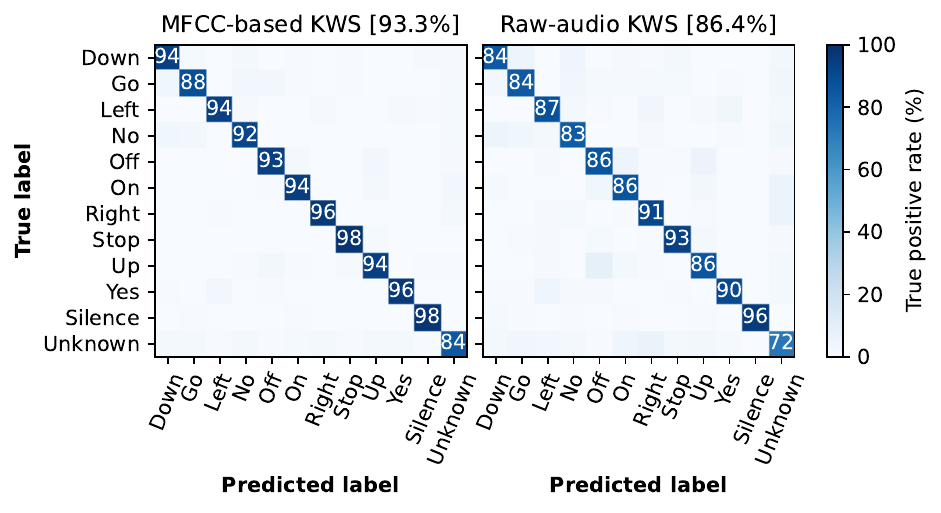}
\caption{Confusion matrices for MFCC-based KWS and raw audio KWS on the 12-class GSCv2 test set, including true positive rates for all keywords.}
\label{fig:confusion_matrix_raw_audio_kws}
\end{figure}

\subsection{Real-time KWS}
\label{sec:real_time_kws}

To characterize both (i) Chameleon’s inference efficiency using its dual-mode \red{multiplier-free} PE array, and (ii) its long-context modeling capabilities using the greedy dilation-aware execution of TCNs, we conduct a series of KWS experiments.

\textit{Benchmarking setup} -- We use the Google Speech Commands V2 (GSCv2) dataset \cite{speech_commands}, which comprises 105,829 utterances from 2,618 speakers across 35 spoken-word classes, with varying sample counts per class. Each sample is recorded for one second at 16 kHz. Ten words serve as command-like keywords: \textit{Yes}, \textit{No}, \textit{Up}, \textit{Down}, \textit{Left}, \textit{Right}, \textit{On}, \textit{Off}, \textit{Stop}, and \textit{Go}. Along with an \textit{unknown command} class (grouping the remaining words) and a \textit{silence} class (one-second audio clips from background noise files), they form the standard 12-way classification setup. We follow the same validation and testing splits as \cite{speech_commands, rybakov2020streaming_kws}. %
To augment the training data, we apply up to \qty{100}{\milli\second} forward or backward shifts (as in \cite{rybakov2020streaming_kws, zhang2017hello_edge}), followed by noise addition with a probability of 0.15. %
Following floating-point training, we perform QAT, during which all augmentations are disabled. For PE array characterization, the audio is transformed into a 28D MFCC feature map \cite{davis1980comparison_mfcc} before it is fed to the DNN. To align with other KWS digital accelerators \cite{vocell, giraldo_efficient_2021, vikram_tinyvers}, we use a window size of \qty{32}{\milli\second} with a \qty{16}{\milli\second} shift for MFCC-extraction, yielding a sequence of 63 timesteps. For long-context modeling, the raw audio data is used.

\textit{Dual-mode PE array characterization} -- To minimize the static-power footprint for real-time MFCC-based KWS, we use a TCN with 16.5k parameters and eight layers, which is small enough to fit in the always-on memories of Chameleon. Running this TCN in the $4\times4$ PE array mode results in a real-time power consumption of \SI{3.1}{\micro\watt} at a clock frequency of \SI{23.3}{\kilo\hertz} with a core voltage of \SI{0.73}{\volt}, achieving a classification accuracy of 93.3\% on the 12-class test set of GSCv2. Executing the same network in the complete $16\times16$ PE array requires a clock frequency of \hltodo{3.67 kHz} to yield a \hltodo{real-time power} of \SI{7.4}{\micro\watt}, with both the MSB memories and the core at \SI{0.73}{\volt}. \Cref{fig:power_comparison_kws} displays the contributions of the core leakage, MSB memories leakage and dynamic power for both real-time KWS scenarios. It can be seen that executing MFCC-based KWS with only the always-on memories leads to a power reduction of 44\% compared to the $16\times16$ baseline. Note also that the dynamic power in the $4\times4$ mode is higher compared to the $16\times16$ mode: since the throughput in the latter is $4\times$ higher, the clock frequency can be downscaled accordingly to achieve the same latency.

\textit{Long-context modeling results} -- In most KWS accelerator designs, the footprint of the MFCC pre-processing exceeds that of the actual DNN inference logic. For example, in the work by Shan~\etal~\cite{nw510kws}, the MFCC extraction accounts for 60\% of the core area, while in \cite{kwantae_kim_23_uw_ring_oscillator_ro_kws} the feature extractor is $\sim\!4\times$ larger than the on-chip DNN accelerator. Similarly, the MFCC extraction module in \cite{vocell} uses the same amount of area as the on-chip DNN accelerator (incurring a $2.4\times$ area overhead when also considering the analog feature extraction). Even fully digital designs like \cite{liu202022nm} allocate 40\% of the core area to MFCC logic. 
To the best of our knowledge, there is currently only one design operating directly on the raw audio sequence without any specific pre-processing logic: the fully analog accelerator in \cite{tan20241_8khz} performs inference on \SI{8}{\kilo\hertz} audio by applying a trained filter bank to \SI{4}{\milli\second} input windows. However, several key DNN processing steps are performed off-chip, preventing end-to-end operation.
In contrast, by enabling long-context inference, our approach eliminates the need for any data-specific pre-processing block, significantly increasing the number of possible deployment scenarios. Using a 24-layer, 118k-parameter model, Chameleon achieves 86.4\% test accuracy on GSCv2 (see \Cref{fig:confusion_matrix_raw_audio_kws}) from \SI{16}{\kilo\hertz} raw audio with \SI{59.4}{\micro\watt} real-time power at a clock frequency of \SI{532}{\kilo\hertz} and a core/MSB memory voltage of \SI{0.73}{\volt} (\autoref{fig:power_comparison_kws}, right).
Although accuracy drops by 7 points, reflecting the challenge of capturing long-range dependencies, it remains competitive, outperforming \cite{kwantae_kim_23_uw_ring_oscillator_ro_kws} with \SI{1.6}{\milli\meter\squared} analog MFCC at 86.0\% 12-class accuracy.

\subsection{Comparison with the State of the Art}
\label{sec:comp_with_sota}

Chameleon is the first SoC supporting both inference and end-to-end on-chip few-shot and continual learning: \autoref{tab:kws_hardware_comparison} compares our design with both KWS and FSL accelerators that have silicon results on GSC and Omniglot datasets, respectively. This comparison highlights the performance of Chameleon across FSL and CL on-device learning, as well as KWS inference scenarios.

\begin{table*}[t]
\caption{Comparison with both KWS accelerators that have reported results on GSC in silicon and with FSL accelerators that have reported results on Omniglot in silicon. The ± shows 95\%
confidence intervals over 100 tasks.}
\label{tab:kws_hardware_comparison}
\centering

\newcommand{\thline}{\specialrule{1.3pt}{0pt}{0pt}}

\begin{adjustbox}{max width=\textwidth}
\begin{threeparttable}

\begin{tabular}{r|ccccccccc}
\thline
 & \multicolumn{4}{c|}{KWS accelerators} & \multicolumn{2}{c|}{CIM accelerators for FSL} & \multicolumn{3}{c}{End-to-end FSL accelerators} \\ \hhline{~|----|--|---}
 & \begin{tabular}[c]{@{}c@{}}Vocell \cite{vocell}\\ JSSC'18\end{tabular} & \begin{tabular}[c]{@{}c@{}}TinyVers \cite{vikram_tinyvers}\\ JSSC'23\end{tabular} & \begin{tabular}[c]{@{}c@{}}Tan~\etal~\cite{tan20241_8khz}\\ JSSC'25\end{tabular} & \begin{tabular}[c]{@{}c@{}}Park~\etal~\cite{park_rebuttal_kws}\\ VLSI'24\end{tabular} & \begin{tabular}[c]{@{}c@{}}Kim~\etal~\cite{other_fsl_cim}\\ TCAS-II'22\end{tabular} & \multicolumn{1}{c|}{\begin{tabular}[c]{@{}c@{}}SAPIENS \cite{sapiens}\\ VLSI'21\end{tabular}} & \begin{tabular}[c]{@{}c@{}}FSL-HDnn \cite{fslhdnn}\\ ESSERC'24\end{tabular} & \multicolumn{2}{c}{\cellcolor[HTML]{EFEFEF}\textbf{This work}} \\ \thline
\begin{tabular}[c]{@{}r@{}}Technology\\ Implementation\\ Core area (\SI{}{\mm\squared})\\ On-chip memory\\ Supply voltage\\ Max. clock frequency\end{tabular} & \begin{tabular}[c]{@{}c@{}}65-nm\\ Digital\\ 2.56\\ 67 kB\\ 0.6-1.2 V\\ 8 MHz\end{tabular} & \begin{tabular}[c]{@{}c@{}}22-nm\\ Digital\\ 6.25\\ 132 kB\\ 0.4-0.9 V\\ 150 MHz\end{tabular} & \begin{tabular}[c]{@{}c@{}}28-nm\\ Analog\\ 0.121\\ 16 kB\\ 0.35/0.9 V\\ 1 MHz\end{tabular} & \multicolumn{1}{c|}{\begin{tabular}[c]{@{}c@{}}65-nm\\ Mixed\\ 1.32\\ 5 kB\\ 0.4-1.2V V\\ 200 kHz\end{tabular}} & \begin{tabular}[c]{@{}c@{}}40-nm LP\\ Mixed signal\\ 0.2\\ 0.8 kB\\ 0.5-1.1 V\\ 80 MHz\end{tabular} & \multicolumn{1}{c|}{\begin{tabular}[c]{@{}c@{}}40-nm\\ Analog\\ 0.0367\\ 8 kB\\ 0.85-1.1 V\\ 200 MHz\end{tabular}} & \begin{tabular}[c]{@{}c@{}}40-nm\\ Digital\\ 11.3\\ 349 kB\\ 0.9-1.2 V\\ 250 MHz\end{tabular} & \multicolumn{2}{c}{\cellcolor[HTML]{EFEFEF}\begin{tabular}[c]{@{}c@{}}40-nm LP\\ Digital\\ 0.74\\ 71 kB\\ 0.6-1.1 V\\ 150 MHz\end{tabular}} \\ \hline
\begin{tabular}[c]{@{}r@{}}Network type\\ Max. \# of on-chip weights\\ Max. demonstrated input length\\ Weight precision\\ Activation precision\end{tabular} & \begin{tabular}[c]{@{}c@{}}LSTM+FC\\ 32k\\ 62\\ 4-bit LUT\\ 8-bit\end{tabular} & \begin{tabular}[c]{@{}c@{}}TCN\\ 400k\\ 60\\ 8-bit\\ 8-bit\end{tabular} & \begin{tabular}[c]{@{}c@{}}FS-CNN\\ 32.8k\\ 8,000\\ 4-bit\\ 3-bit unsigned\end{tabular} & \multicolumn{1}{c|}{\begin{tabular}[c]{@{}c@{}}TENet\\ 7.4k\\ 46\\ 4-bit $\log_2$\\ 4-bit unsigned\end{tabular}} & \multicolumn{2}{c|}{\begin{tabular}[c]{@{}c@{}}\;\\ \textcolor{bred}{Distance computation only,}\\ \textcolor{bred}{off-chip FP32 embedder}\\ \;\end{tabular}} & \begin{tabular}[c]{@{}c@{}}CNN\\ 2.1k\\ 28\\ Bfloat-16\\ 16-bit$^5$\end{tabular} & \multicolumn{2}{c}{\cellcolor[HTML]{EFEFEF}\begin{tabular}[c]{@{}c@{}}TCN+FCs\\ 133k\\ \textbf{16,000}\\ 4-bit $\log_2$\\ 4-bit unsigned\end{tabular}} \\ \hline
\begin{tabular}[c]{@{}r@{}}Inference support\\ End-to-end inference\\ Full on-chip weight storage\\ FSL support\\ End-to-end FSL\\ CL support\end{tabular} & \begin{tabular}[c]{@{}c@{}}\cmark\\ \cmark\\ \cmark\\ \textcolor{bred}{\xmark}\\ \textcolor{bred}{\xmark}\\ \textcolor{bred}{\xmark}\end{tabular} & \begin{tabular}[c]{@{}c@{}}\cmark\\ \cmark\\ \cmark\\ \textcolor{bred}{\xmark}\\ \textcolor{bred}{\xmark}\\ \textcolor{bred}{\xmark}\end{tabular} & \begin{tabular}[c]{@{}c@{}}\cmark\\ \textcolor{bred}{\xmark}\\ \cmark\\ \textcolor{bred}{\xmark}\\ \textcolor{bred}{\xmark}\\ \textcolor{bred}{\xmark}\end{tabular} & \multicolumn{1}{c|}{\begin{tabular}[c]{@{}c@{}}\cmark\\ \textcolor{bred}{\xmark}\\ \cmark\\ \textcolor{bred}{\xmark}\\ \textcolor{bred}{\xmark}\\ \textcolor{bred}{\xmark}\end{tabular}} & \multicolumn{2}{c|}{\begin{tabular}[c]{@{}c@{}}\textcolor{bred}{\xmark \; (off-chip embedder)}\\ \textcolor{bred}{\xmark}\\ \textcolor{bred}{\xmark}\\ \cmark\\ \textcolor{bred}{\xmark}\\ \textcolor{bred}{\xmark}\end{tabular}} & \begin{tabular}[c]{@{}c@{}}\cmark\\ \textcolor{bred}{\xmark}\\ \textcolor{bred}{\xmark}\\ \cmark\\ \textcolor{bred}{\xmark}\\ \textcolor{bred}{\xmark}\end{tabular} & \multicolumn{2}{c}{\cellcolor[HTML]{EFEFEF}\begin{tabular}[c]{@{}c@{}}\cmark\\ \cmark\\ \cmark\\ \cmark\\ \cmark\\ \cmark\end{tabular}} \\ \hline
\begin{tabular}[c]{@{}r@{}}FSL method / distance metric\\ FSL embedding dimension\\ \# of FSL classes\\ \# of FSL shots\end{tabular} & \multicolumn{4}{c|}{\begin{tabular}[c]{@{}c@{}}\;\\ \textcolor{bred}{Do not support FSL}\\ \;\end{tabular}} & \begin{tabular}[c]{@{}c@{}}MANNs, L1\\ 64\\ 1-25\\ 1-5\end{tabular} & \multicolumn{1}{c|}{\begin{tabular}[c]{@{}c@{}}MANNs, L1\\ 32\\ 32\\ 1\end{tabular}} & \begin{tabular}[c]{@{}c@{}}HDC, Hamming\\ 16-1024\\ 2-128\\ 1, 5$^5$ \end{tabular} & \multicolumn{2}{c}{\cellcolor[HTML]{EFEFEF}\begin{tabular}[c]{@{}c@{}}PNs, $\text{L2}^2$\\ \textbf{1-1024}\\ \textbf{1-256}\\ \textbf{1-128}\end{tabular}} \\ \thline
\begin{tabular}[c]{@{}r@{}}\textbf{Omniglot FSL/CL}\\ Model size\\ End-to-end power\\ End-to-end latency\end{tabular} & \multicolumn{4}{c|}{\begin{tabular}[c]{@{}c@{}}\;\\ \textcolor{bred}{Do not support FSL}\\ \;\end{tabular}} & \begin{tabular}[c]{@{}c@{}}\;\\ 7.46 MB\\ N/A\\ N/A\end{tabular} & \multicolumn{1}{c|}{\begin{tabular}[c]{@{}c@{}}\;\\ 447 kB\\ N/A\\ N/A\end{tabular}} & \begin{tabular}[c]{@{}c@{}}(@ 100 MHz)\\ 5.5 MB$^5$\\ 27 mW\\ 53 ms\end{tabular} & \multicolumn{2}{c}{\cellcolor[HTML]{EFEFEF}\begin{tabular}[c]{@{}c@{}}(@ 100 kHz)\;\;\;\;\;(@ 100 MHz)\\ \textbf{59 kB}\\ \textbf{12.9 \textmu W}\;\;\;\;\;\;\;\;\;\;\;\;\textbf{11.6 mW}\\ \;0.54s\;\;\;\;\;\;\;\;\;\;\;\;\;\;\;\textbf{0.59 ms}\end{tabular}} \\ \hline
\begin{tabular}[c]{@{}r@{}}\textbf{Omniglot FSL accuracy}\\ 5-way (1, 5 shots)\\ 20-way (1, 5-shots)\\ 32-way, 1-shot\end{tabular} & \multicolumn{4}{c|}{\begin{tabular}[c]{@{}c@{}}\;\\ \textcolor{bred}{Do not support FSL}\\ \;\end{tabular}} & \begin{tabular}[c]{@{}c@{}}\;\\ 93.4\%, 98.3\%\\  -\\ -\end{tabular} & \multicolumn{1}{c|}{\begin{tabular}[c]{@{}c@{}}\;\\ -\\ -\\ 72\%\end{tabular}} & \begin{tabular}[c]{@{}c@{}}\;\\ 79.0\%,\;\;\; -\;\;\;\;\\ \;\;\;\;\;\;-\;\;\;\;, 79.5\%\\ -\end{tabular} & \multicolumn{2}{c}{\cellcolor[HTML]{EFEFEF}\begin{tabular}[c]{@{}c@{}}\;\\ \textbf{\omnfwayoshot}, \textbf{\omnfwayfshot}\\ \textbf{\omntwayoshot}, \textbf{\omntwayfshot}\\ \textbf{\omnthwayoshot}\end{tabular}} \\ \hline
\begin{tabular}[c]{@{}r@{}}\textbf{Omniglot 250-way CL acc.}\\ 1-shot (final, avg.)\\ 2-shot (final, avg.)\\ 5-shot (final, avg.)\\ 10-shot (final, avg.)\end{tabular} & \multicolumn{7}{c}{\begin{tabular}[c]{@{}c@{}}\;\\ \textcolor{bred}{Do not support continual learning}\\ \;\end{tabular}} & \multicolumn{2}{c}{\cellcolor[HTML]{EFEFEF}\begin{tabular}[c]{@{}c@{}}\;\\ \textbf{57.6 ± 1.0\%}, \textbf{70.3\%}\\ \textbf{69.2 ± 0.6\%}, \textbf{80.2\%}\\ \textbf{79.3 ± 0.7\%}, \textbf{87.1\%}\\ \textbf{82.2 ± 0.4\%}, \textbf{89.0\%}\end{tabular}} \\ \thline
\begin{tabular}[c]{@{}r@{}}\textbf{GSC 12-class KWS}\\ Pre-processing\\ Accuracy @ dataset version\\ Latency\\ Real-time power\\ Model size\end{tabular} & \begin{tabular}[c]{@{}c@{}}(@ 250kHz)\\ MFCC\\ 90.87\% (v1)\\ 16 ms\\ 10.6 \textmu W$^1$\\ 16 kB\end{tabular} & \begin{tabular}[c]{@{}c@{}}(@ 5 MHz)\\ MFCC\\ 93.3\% (v1)\\ 11 ms\\ 193 \textmu W\\ 23 kB\end{tabular} & \begin{tabular}[c]{@{}c@{}}(@ 1 MHz)\\ None\\ 91.8\% (v2)\\ 2 ms\\ \textbf{1.73 \textmu W}\\ 11 kB\end{tabular} & \multicolumn{1}{c|}{\begin{tabular}[c]{@{}c@{}}(@ 200 kHz)\\ MFCC\\ 92.7\% \\ 10.7 ms\\ 4.68 \textmu W$^6$\\ \textbf{3.7 kB}\end{tabular}} & \multicolumn{3}{c}{\begin{tabular}[c]{@{}c@{}}\;\\ \;\\ \textcolor{bred}{Do not support inference on temporal data}\\ \;\\ \;\end{tabular}} & \cellcolor[HTML]{EFEFEF}\begin{tabular}[c]{@{}c@{}}(@ \hltodo{23.3 kHz})\\ MFCC\\ \textbf{93.3\%} (v2)\\ \hltodo{16} ms\\ 3.1 \textmu W\\ 8.5 kB\end{tabular} & \cellcolor[HTML]{EFEFEF}\begin{tabular}[c]{@{}c@{}}(@ 532 kHz)\\ None\\ 86.4\% (v2)\\ \hltodo{63}~\SI{}{\micro\second}\\ 59.4 \textmu W\\ 60 kB\end{tabular} \\ \thline
\begin{tabular}[c]{@{}r@{}}Peak GOPS\\ Peak TOPS/W\end{tabular} & \begin{tabular}[c]{@{}c@{}}0.13\\ 0.45$^5$\end{tabular} & \begin{tabular}[c]{@{}c@{}}17.6\\ 17\end{tabular} & \begin{tabular}[c]{@{}c@{}} N/C\\ N/C \end{tabular} & \multicolumn{1}{c|}{\begin{tabular}[c]{@{}c@{}} 0.0144\\ 3.1 \end{tabular}} & N/A$^3$ & \multicolumn{1}{c|}{N/A$^4$} & \begin{tabular}[c]{@{}c@{}}\textbf{154}\\ 5.7\end{tabular} & \multicolumn{2}{c}{\cellcolor[HTML]{EFEFEF}\begin{tabular}[c]{@{}c@{}}76.8\\ \textbf{6.0}\end{tabular}} \\ \thline
\end{tabular}

\begin{tablenotes}
  \item[1] Excludes on-chip MFCC computation power. $^2$~Interpolated metrics. $^3$~Reported metrics are for distance computation only and amount to 4.85$^5$ GOPS and 27.7 TOPS/W. $^4$~Reported metrics are for distance computation only and amount to 0.4 GOPS$^5$ and 0.118 TOPS/W. $^5$~Estimated metrics. $^6$ Excludes analog front-end power.
  \end{tablenotes}
\end{threeparttable}
\end{adjustbox}

\end{table*}

\textit{FSL performance} -- Chameleon is the only end-to-end FSL chip that supports sequential data, while offering fully flexible support for embedding dimensions, number of ways, and shots. \red{Chameleon is also the first work that not only enables multiplier-free inference but also completely multiplier-free FSL, without requiring any modifications of the inference datapath.} Furthermore, across all FSL evaluation scenarios, Chameleon sets a new record, despite using an embedding DNN that is $7.6$–$126\times$ smaller than other FSL chips. %
Comparing with CIM accelerators for FSL distance computation, Kim~\etal~\cite{other_fsl_cim} implement MANNs (\autoref{sec:fsl_on_chip_suitability}) with an off-chip FP32 embedder of \SI{7.46}{\mega\byte}, resulting in 3.4\% and 0.5\% lower accuracy than Chameleon for the 5-way 1- and 5-shot cases, respectively. SAPIENS \cite{sapiens} employs an RRAM-based backend to perform FSL, also using MANNs. On 32-way 1-shot Omniglot, SAPIENS achieves 72\% accuracy using an FP32-based 4-layer CNN of \SI{447}{\kilo\byte}, which is 11.3\% lower than Chameleon using an $8\times$ larger network. %
To the best of our knowledge, the only other FSL accelerator with an on-chip embedder reporting metrics on Omniglot is FSL-HDnn~\cite{fslhdnn}, which uses HDC for FSL. It employs a Bfloat-16-based \SI{5.5}{\mega\byte} ResNet-18 model for generating embeddings, achieving \SI{79}{\percent} accuracy for 5-way 1-shot and \SI{79.5}{\percent} for 20-way 5-shot tasks on Omniglot. These accuracies are significantly lower than Chameleon, which uses a network $\sim\!\!100\times$ smaller that fits fully on-chip. Indeed, even though FSL-HDnn has a core area $12\times$ larger than Chameleon, only 0.07\% of its DNN weights can be stored on-chip at any time. 
Furthermore, while FSL-HDnn uses distance-based classification, similar to Chameleon, its HDC encoding adds 120\% power and 26\% area overhead. Chameleon fully sidesteps this costly HDC encoding by computing distances directly on the embeddings. Overall, Chameleon uses $2.5\times$ lower power than FSL-HDnn at the same clock frequency, while reducing the end-to-end latency by $90\times$, yielding an energy per shot $210\times$ lower.

\textit{CL performance} -- Chameleon is the first design to offer scalable end-to-end CL fully on-chip, establishing the first baseline for 250-way CL.
Its unified learning and inference architecture, which enables a minimum CL power consumption of \SI{12.9}{\micro\watt}, allows limiting the memory overhead of CL to scale with only 26 bytes per way on Omniglot (\Cref{eq:l2_is_linearshifted}). This is a negligible 0.04\% of the total DNN footprint.
Hence, by reusing its DNN weight memory to store FC parameters for new classes, Chameleon flexibly repurposes on-chip memory to scale beyond prior class limits: even with 90\% of memory allocated to the deployed TCN, it can still accommodate 250 learned Omniglot classes. 
In contrast, FSL CIMs like Kim~\etal~\cite{other_fsl_cim} and SAPIENS \cite{sapiens} are limited to 25 and 32 classes, respectively, due to their fixed array sizes. FSL-HDnn~\cite{fslhdnn} also uses a fixed separate memory to store its high-dimensional encodings, which is 80\% larger than all on-chip memory in Chameleon. Yet, only up to 20-way learning is demonstrated, scaling with \SI{8.2}{\kilo\byte}/way on Omniglot.

\textit{Real-time KWS performance} – Chameleon matches SotA 93.3\% accuracy on GSC using MFCCs, with the lowest reported real-time end-to-end power (\SI{3.1}{\micro\watt}) and the highest peak throughput (76.8~GOPS) among end-to-end KWS inference accelerators. Compared to Vocell~\cite{vocell}, Chameleon offers 2.5\% higher accuracy with $2\times$ lower power and $600\times$ higher peak GOPS. Versus TinyVers~\cite{vikram_tinyvers}, which uses a TCN on a RISC-V SoC with a reconfigurable accelerator, Chameleon matches accuracy with a $3\times$ smaller model, and $64\times$ lower power. Tan~\etal~\cite{tan20241_8khz} avoid MFCC pre-processing to infer on raw audio instead: using a fully analog design, they improve raw-audio performance by 5.4 accuracy points at $1.8\times$ lower power, with an estimated $25\times$ higher peak efficiency compared to Chameleon. However, it does not perform end-to-end on-chip inference as several key steps of inference computation (\textit{e.g.}, ReLU activation, BN, stride adjustments, max pooling, activation clipping) are carried out off-chip.
\red{The design by Park~\etal~\cite{park_rebuttal_kws}, on the other hand, does perform end-to-end KWS using the smallest total SRAM footprint across the compared designs of only 5 kB. However, the design's real-time power consumption is about 50\% higher at a 0.6\%-point reduction in accuracy compared to Chameleon, while also relying on an embedded MFCC feature extractor for inference, indicating the inability for the neural network to operate directly on the raw data.}

\section{Conclusion}
\label{sec:conclusion}

\vspace{-1pt}

In this work, we presented Chameleon, the first scalable SoC to enable end-to-end on-chip FSL and few-shot CL on sequential data, without compromising inference accuracy or efficiency. We achieve this through three key contributions.

First, by integrating FSL and CL into the inference process, we propose a unified learning and inference architecture built on PNs~\cite{snell2017prototypical}. %
The learning integration incurs only $<0.02$\% latency and 0.5\% core area overhead. Second, our greedy dilation-aware TCN execution enables the first \SI{16}{\kilo\hertz} raw audio KWS on GSCv2, achieving a 160$\times$ larger receptive field than prior end-to-end KWS accelerators using only \SI{2}{\kilo\byte} activation memory -- 4$\times$ less than SotA TCN accelerators, while avoiding a 60-400\% area increase for audio-specific pre-processing blocks. Third, by using $\log_2$ quantization, we enable a \red{multiplier-free} PE array. Combined with its dual-mode operation through system-level power gating, this yields a peak throughput in $16\times16$-mode that is $4.3\times$ higher than previous KWS accelerators or $2\times$ lower real-time power in $4\times4$-mode.

Chameleon establishes new accuracy records for end-to-end on-chip FSL on the Omniglot dataset (\omnfwayoshot\;5-way 1-shot, \omntwayfshot\;20-way 5-shot) at $2.5\times$ lower power than previous on-chip embedder FSL work \cite{fslhdnn}. It is the only end-to-end FSL accelerator with all weights on-chip and enables FSL starting at \SI{12.9}{\micro\watt}.
Considering CL, although several designs exist, none offer a fully end-to-end SoC solution with comparable accuracy, footprint, and scalability.
Hence, Chameleon establishes the first baseline for end-to-end on-chip CL on Omniglot, demonstrating 250-way few-shot CL (82.2\% final accuracy for 10 shots). Beyond this learning performance, Chameleon maintains a SotA accuracy of 93.3\% for 12-class GSCv2 compared to inference-only KWS accelerators. 

\red{Ultimately, by enabling (i)~a unified learning and inference architecture, (ii)~a flexible PE-array architecture that allows multiplier-free learning with low-resolution quantization levels that are typical of inference-only accelerators, and (iii)~an efficient TCN embedder that allows scaling this approach to long-range temporal dependencies up to raw-audio processing}, Chameleon paves the way for efficient lifelong learning on extreme-edge devices.

\section*{Acknowledgement}

The authors thank Prof. Kofi Makinwa, Nicolas Chauvaux, Dr. Martin Lefebvre, and Dr. Adrian Kneip for their feedback, Dr. Filipe Cardoso for dicing and Nuriel Rozsa for die photos. This publication was partly financed by the Dutch Research Council (NWO) as part of the project AdaptEdge with file number 20267 in the NWO Talent Programme -- Veni.

\printbibliography
\begin{IEEEbiography}[{\includegraphics[width=1in,height=1.3in,clip,keepaspectratio]{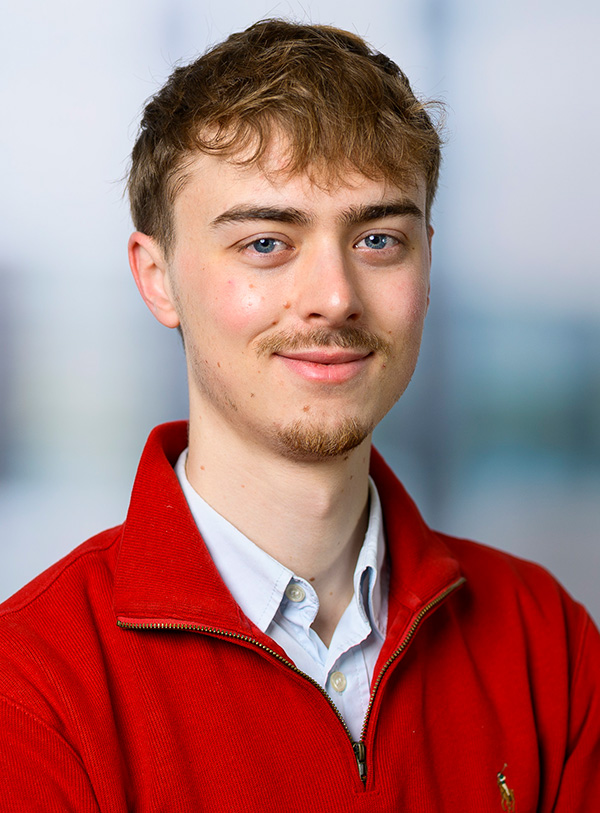}}]{Douwe den Blanken} (Graduate Student Member, IEEE) received the M.Sc. degree (\textit{with honors}) in embedded systems from the Delft University of Technology (TU Delft), Delft, The Netherlands, in 2023, where he is currently pursuing the Ph.D. degree, under the supervision of Prof. C. Frenkel. His current research interests include efficient learning algorithms and their implementation in silicon, as well as the quantization and acceleration of modern DNNs. 
\end{IEEEbiography}
\vfill
\newpage

\begin{IEEEbiography}[{\includegraphics[width=1.25in,height=1.3in,clip,keepaspectratio]{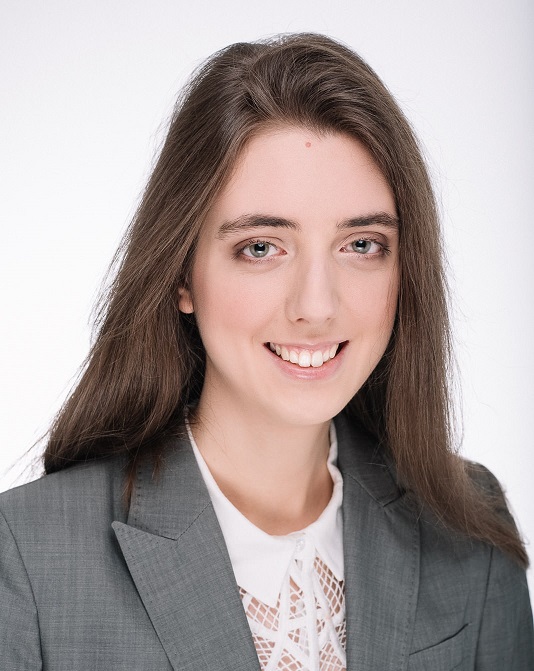}}]{Charlotte Frenkel}
(Member, IEEE) received the M.Sc. degree (\textit{summa cum laude}) in Electromechanical Engineering and the Ph.D. degree in Engineering Science from Universit\'e catholique de Louvain (UCLouvain), Louvain-la-Neuve, Belgium in 2015 and 2020, respectively. In February 2020, she joined the Institute of Neuroinformatics, UZH and ETH Zurich, Switzerland, as a postdoctoral researcher. She is an Assistant Professor at Delft University of Technology, Delft, The Netherlands, since July 2022, and holds a Visiting Faculty Researcher position with Google since October 2024.

Her research aims at bridging the bottom-up (bio-inspired) and top-down (engineering-driven) design approaches toward neuromorphic intelligence, with a focus on hardware-algorithm co-design for (Neuro)AI, digital hardware accelerators, and brain-inspired on-device learning.

Dr. Frenkel received a best paper award at the IEEE International Symposium on Circuits and Systems (ISCAS) 2020 conference in the \textit{Neural Networks} track, and her Ph.D. thesis was awarded the FNRS-FWO / Nokia Bell Scientific Award 2021 and the FNRS-FWO / IBM Innovation Award 2021. In 2023, she was awarded prestigious Veni and AiNed Fellowship grants from the Dutch Research Council (NWO). She presented several invited talks, including keynotes at the tinyML EMEA technical forum 2021 and at the Neuro-Inspired Computational Elements (NICE) neuromorphic conference 2021. She serves or has served as a program co-chair of NICE 2023-2024 and of the tinyML Research Symposium 2024, as a co-lead of the NeuroBench initiative for benchmarks in neuromorphic computing since 2022, as a TPC member of IEEE ESSERC for 2022-2024, and as an associate editor for the IEEE Transactions on Biomedical Circuits and Systems since 2022.
\end{IEEEbiography}

\vfill

\end{document}